\documentclass[preprint2]{aastex}  

\shorttitle{THE SHOCK-INDUCED STAR FORMATION SEQUENCE}
\shortauthors{Mart\'{\i}nez-Garc\'ia et al.}

\usepackage{mathrsfs} 
\newcommand{\jtwo}{\mbox{$2\rightarrow1$}}
\newcommand{\jone}{\mbox{$1\rightarrow0$}}

\begin{document}

\title{THE SHOCK-INDUCED STAR FORMATION SEQUENCE RESULTING FROM A CONSTANT SPIRAL PATTERN SPEED}

\author{Eric E. Mart\'inez-Garc\'ia\altaffilmark{1} \& Iv\^anio Puerari\altaffilmark{1}}
\affil{1 Instituto Nacional de Astrof\'isica, \'Optica y Electr\'onica (INAOE), Aptdo. Postal 51 y 216, 72000 Puebla, Pue., M\'exico.\\
\email{ericmartinez@inaoep.mx, puerari@inaoep.mx}}

\begin{abstract}

We utilize a suite of multiwavelength data, of 9 nearby spirals,
to analyze the shock-induced star formation sequence, that may result
from a constant spiral pattern speed.
The sequence involves tracers as the H{\rm{I}}, CO, $24\micron$, and FUV,
where the spiral arms were analyzed with Fourier techniques
in order to obtain their azimuthal phases as a function of radius.
It was found that only two of the objects, NGC~628 and NGC~5194,
present coherent phases resembling the theoretical expectations,
as indicated by the phase shifts of CO-$24\micron$.
The evidence is more clear for NGC~5194, and moderate for NGC~628.
It was also found that the phase shifts are different for the two
spiral arms. With the exception on NGC~3627, a two-dimensional Fourier analysis
showed that the rest of the objects do not exhibit bi-symmetric spiral
structures of stellar mass, i.e., grand design spirals.
A phase order inversion indicates a corotation radius of $\sim89\arcsec$
for NGC~628, and $\sim202\arcsec$ for NGC~5194.
For these two objects, the CO-H$\alpha$ phase shifts corroborate
the CO-$24\micron$ azimuthal offsets. Also for NGC~5194, the CO-$70\micron$, CO-$140\micron$,
and CO-$250\micron$ phase shifts indicate a corotation region.

\end{abstract}


\section{Introduction}

Nowadays, the majority of numerical simulations show no agreement for the hypotheses of long-lived
quasi-steady spiral structure~\citep{lin64,ber89}. Instead, recurrent short-lived transient spirals are often 
obtained~\citep[e.g.,][]{sel11,wada11,fuj11,gra12,bab13,don13,roc13}.
However, azimuthal age gradients studies suggest otherwise~\citep[e.g.,][]{mar09a,mar13},
and are in agreement with other observational studies
of age patterns (or offsets) across spiral arms~\citep[e.g.,][]{efr85,egu09,gro09,san11,ced13}.
In a study of the age and the star formation rate of HII regions,~\citet{ced13}, had found newly evidence
of the triggering of star formation by density waves (DWs).
Also, the studies by ~\citet{sca11}, and~\citet{sca13},
related to the ``breaks'' in the radial metallicity distribution found near the corotation,
imply that spiral arms may be long-lived structures.
Any discontinuities in the radial metallicity profiles would be smoothed out
unless a dominant pattern speed exists with a sole corotation~\citep{sca13}.

Altogether these observations suggest the presence of spiral DWs
in galaxies. Age/color gradients, across spiral arms, indicate that
the pattern speed for these objects must be a constant for all radii.
From this point it is inferred that the patterns are long-lived
as proposed by density wave (DW) theory. Nevertheless, there is no quantification
of the lifetime of the spirals.
However, it is known that spirals are expected to appear at redshifts of $z\sim1-2$~\citep{elm14}.
Even more, an apparent spiral structure has been found at a redshift of $z=2.18$~\citep{law12}.
For those spirals to survive at $z=0$ would require lifetimes of at
least $\sim8$~Gyr. Whether these structures are long-lasting patterns
is still an unresolved issue.

The DW theory is not the only one that predicts a constant angular speed for the spiral pattern.
The ``manifold theory''~\citep{rom06,vog06a,vog06b,vog06c,rom07,tso08,tso09,atha09a,atha09b,atha10},
where chaotic orbits play a major role in generating spiral structure,
also predicts the same behavior. In the view of~\citet{rom06,rom07,atha09a,atha09b,atha10},
the ``manifolds'' behave as tube structures that trap
chaotic orbits within them. In contrast to DW theory, where the stars that make up the spirals do not remain
within them, the manifold theory (in the view of Athanassoula et al.) 
predicts that stars must follow along the arms in a direction away
from corotation. Manifold theories had only been proposed for barred-spiral systems,
where the bar's gravitational potential generates unstable Lagrange points near corotation.
A manifold theory to explain spiral structure in normal (or weakly barred) galaxies
is under development (Athanassoula 2013, private communication).

Regardless of the theory involved, or the lifetime of the patterns,
many of the observational evidence (discussed above) suggests that
some disk galaxies must present a fixed spiral pattern speed for all radii at the present moment.
If such spiral arms trigger star formation (SF), besides the previously discussed azimuthal age/color gradients,
some observational tracers for different stages of the star formation
sequence should show a spatial ordering~\citep[e.g.,][]{foy11}. From upstream to downstream before the corotation,
and in the corotating frame, we may have dense H{\rm{I}} tracing
the compressed gas, CO tracing molecular gas, $24\micron$ emission tracing dust-obscured star formation~\citep{cal07},
and UV emission tracing unobscured young stars. This research involves the analysis of spiral galaxies in search for this
sequence of stages of star formation. For this purpose we chose the Fourier method of~\citet{pue97},
that determines the phases of the intensities for spiral arms as observed in different wavelengths.

\section{Data sample}

This research utilizes common data of nearby spirals in 
SINGS~\citep{ken03}, THINGS
\citep{wal08}, HERACLES~\citep{ler09}, and GALEX~\citep{gil04,gil07}.
The Spitzer Infrared Nearby Galaxies Survey (SINGS) is a Legacy survey
aimed to characterize the infrared emission across the entire range of galaxy properties and star formation environments.
SINGS integrates visible/UV and IR/submillimeter studies into a coherent self-consistent whole.
The H{\rm{I} Nearby Galaxy Survey (THINGS) was a program undertaken at the National Radio Astronomy Observatory (NRAO), Very Large Array (VLA),
to perform 21-cm H{\rm{I} observations of nearby galaxies. The goal of THINGS was to investigate galaxy morphology, star formation and mass
distribution across the Hubble sequence. Data from THINGS complement SINGS.
The HERA CO-Line Extragalactic Survey (HERACLES) used the IRAM 30-m telescope to map CO emission from nearby galaxies. 
HERACLES was built to complement THINGS, SINGS, and associated surveys.
The GALEX Nearby Galaxies Survey (NGS) intends to answer fundamental open questions on galaxy evolution
and UV properties of galaxies. The NGS sample was partially built using the Spitzer's Reserved Observations Catalog.
The NGS was the forerunner of the GALEX Ultraviolet Atlas of Nearby Galaxies~\citep{gil07}.
Together these surveys give us a bolometric view of local galaxies in the universe.

Our sample consists of 9 spiral galaxies with data from 
$24\micron$~\citep[SINGS,][]{ken03}, 
H{\rm{I}}~\citep[THINGS,][]{wal08},
$^{12}$CO~$J=\jtwo$~\citep[HERACLES,][]{ler09},
and FUV\footnote{
With the exception of NGC~3521 where we use the
NUV~($\lambda_{\mathrm{eff}}=2267\AA$) waveband.}
\citep[GALEX, $\lambda_{\mathrm{eff}}=1516\AA$,][]{gil04,gil07}.
The sample was chosen following~\citet{foy11} who used a
polar cross-correlation method and found little evidence of offsets
for the different star formation stages assuming a DW scenario.
Here we examine 9 objects of their sample adopting a different approach (phases via Fourier analysis, see section~\ref{analysis}).
We ensure that our selected objects have data in each of the frequencies of interest.
The objects are shown in figure~\ref{fig_sample}.
Each frame was deprojected with the parameters presented in table~\ref{tbl-1}.
The images were registered using the WCS information of the THINGS data.
The H{\rm{I}, $24\micron$, and FUV data were convolved with a circularly symmetric Gaussian function with
a FWHM of $13\arcsec$, this corresponds to the resolution of the HERACLES data~\citep{ler09}.

The morphological classification~\citep[from RC3,][see table~\ref{tbl-1}]{dva91} indicates 
that four of our objects (NGC 628, NGC 2841, NGC 5055, and NGC 5194) have no evident bar in general (SA galaxies),
other four (NGC 2403, NGC 3521, NGC 3627, and NGC 6946) have characteristics intermediate between
barred and nonbarred galaxies (SAB galaxies), and one object (NGC 3351)
has a clear and well-defined bar (SB galaxy).
The arm classification~\citep{elm82,elm87} indicates that five objects (NGC 628, NGC 3351,
NGC 3627, NGC 5194, and NGC 6946) have two symmetric arms, i.e. arm class greater than 5, in which only NGC 5194 has two
long symmetric arms dominating the optical disk, i.e. arm class 12.
Regarding the environmental information of our sample of galaxies (cf. NED\footnote{http://ned.ipac.caltech.edu/}),
NGC 628 is a X-ray-faint group member~\citep{sen06}; 
NGC 5194 is a Pair member~\citep{kar72};
NGC 2403, NGC 2841, NGC 3521, and NGC 6946 are isolated galaxies~\citep{kar73,san75};
while NGC 3351, NGC 3627, and NGC 5055 are in a radial-velocity based grouping~\citep{mah98}.


\begin{deluxetable}{llccccc}
\tabletypesize{\scriptsize}
\tablecaption{Parameters of the data~\label{tbl-1}}
\tablewidth{0pt}
\tablehead{
\colhead{Object} &
\colhead{Type} &
\colhead{AC} &
\colhead{Incl.~($\degr$)} &
\colhead{P.A.~($\degr$)} &
\colhead{$v_{\mathrm{rot}}$(km s$^{-1})$} &
\colhead{Dist.(Mpc)}
}
\startdata

NGC~628  (M74)  & SA(s)c        &  9 &  7       &  20       & 179       &  9.93 \\ 
NGC~2403        & SAB(s)cd      &  4 & 63       & 124       & 118       &  4.55 \\
NGC~2841        & SA(r)b        &  3 & 63       & 148       & 327       &  12.26 \\
NGC~3351 (M95)  & SB(r)b        &  6 & 41       & 192       & 187       &  9.00 \\
NGC~3521        & SABrs)bc      &  3 & 65       & 162       & 234       &  8.00 \\
NGC~3627 (M66)  & SAB(s)b       &  7 & 62       & 173       & 183       &  6.53 \\
NGC~5055 (M63)  & SA(rs)bc      &  3 & 59       & 102       & 208       &  8.33 \\
NGC~5194 (M51)  & SA(s)bc       & 12 & 20       & 172       & 211       &  9.12 \\
NGC~6946        & SAB(rs)cd     &  9 & 33       & 242       & 181       &  5.49 \\

\enddata

\tablecomments{
Column 2: morphological type~\citep[RC3,][]{dva91}.
Column 3: arm class~\citep{elm82,elm87}.
Columns 4 and 5: inclination and position angles in degrees~\citep{foy11}.
Column 6: observed maximum velocity of rotation~\citep{pat03}, corrected for inclination (column 4).
Column 7: Hubble flow distance from the NASA/IPAC extragalactic database (Virgo + Great Attractor + Shapley Supercluster),
$H_{0}=73$ km/s/Mpc, $\Omega_{\mathrm{matter}}=0.27$, and $\Omega_{\mathrm{vacuum}}=0.73$.
}

\end{deluxetable}


\section{Analysis}~\label{analysis}

Our adopted method is based on the Fourier analysis of azimuthal intensities
to locate the corotation radius ($R_{CR}$) in spiral galaxies~\citep{pue97}.
The Fourier transform

\begin{equation}
    \mathscr{F}_{m}(R)=\int_{-\pi}^{\pi} I_{R}(\theta) e^{-im \theta} d\theta,
\end{equation}

\noindent where $I_{R}$ is the intensity of radiation as a function of radius ($R$),
is computed for the object's image at a certain waveband.
The azimuthal phase, $\Theta(R)$, is then computed as

\begin{equation}
    \Theta(R) = \tan^{-1} \left\{ \frac{\mathrm{Re}[\mathscr{F}_{m}(R)]}{\mathrm{Im}[\mathscr{F}_{m}(R)]} \right\}.    
\end{equation}

For bi-symmetrical spirals we have $m=2$, then $\mathscr{F}_{m}(R) = \mathscr{F}_{2}(R)$.
This method has the advantage of getting the phase of the $m=2$ spiral even
in the presence of noise or when the data is mixed with radiation from other sources (e.g., foreground and background objects).
For the case when the spiral pattern speed is a constant for all radii (as in the DW scenario)
we expect the H{\rm{I}},\footnote{We notice that we keep the H{\rm{I}} at the beginning of the sequence,
assuming that it is tracing the highly compressed gas, despite the fact that it is affected by the process of molecular
photodissociation in star-forming regions~\citep[e.g.,][]{all86,ran92,lou13}.}
CO, $24\micron$, and FUV spirals to present Fourier phases in a sequential order.
This order should be inverted near corotation, as shown in figure~\ref{fig_shifts}.\footnote{
As compared with the phases, the maximums of the azimuthal emission in the arms would be displaying a reverse order.
Also the curvature of the lines plotted in panel ``a'' of figure~\ref{fig_shifts} changes from concave to convex if 
the maximum of the azimuthal emission is graphed instead of the respective phases.}
This is because before corotation the angular velocity of the disk material,
$\Omega(R)$, is higher than the angular velocity of the spiral pattern, $\Omega_{p}$.
The angular velocities are equal at corotation, and beyond corotation
$\Omega_{p}$ exceeds $\Omega(R)$.

The curves for the phases in figure~\ref{fig_shifts} can be described with the equation

\begin{equation}~\label{eqTHETA}
  \Theta(R) = \frac{\ln{R}}{\tan{(P_{\mathrm{shock}})}} + \Theta_{0} - t_{\mathrm{SF}}\left(\frac{v_{\mathrm{rot}}}{R} - \Omega_{p} \right),  
\end{equation}

\noindent where $P_{\mathrm{shock}}$ is the pitch angle of the spiral shock, $\Theta_{0}$ is a constant
that determines the angular spatial location of the spiral shock (in the corotating frame), 
$t_{\mathrm{SF}}$ is the time elapsed between the shock and the corresponding star formation stage,
and $v_{\mathrm{rot}}$ is the circular velocity of rotation. 
In this sense, each phase angle curve dependence on the waveband is given by the quantity $t_{\mathrm{SF}}$.
We must remark that equation~\ref{eqTHETA} assumes circular orbits for
the gas and stars involved in the star formation sequence. However,
after passing through a spiral shock
the material flows to some extent along the arms, and then flows to the interam region~\citep{mar09b,dob10}.
\citet{mar09b} has shown that by assuming a circular motion model, the determination of the pattern speed ($\Omega_{p}$)
is affected for radii different from corotation ($R_{CR}$), and that this effect does not prevents that star formation offsets
(or the corresponding azimuthal age gradients) are observed across the arms. The material's non-circular motions near
the spiral arms (or streaming motions) can distort the curves shown in figure~\ref{fig_shifts},
but only for radii away from $R_{CR}$.

The H{\rm{I}}, CO, $24\micron$, and FUV data for our sample of objects were analyzed via the Fourier phases method.
We assumed trailing spirals and adopt positive values for the phases in the direction of rotation.

\subsection{Mass spirals of the grand design type}~\label{mass_spirals}

Taking into account the structural type of their arms
only {\it{grand design}} spirals are suitable to
test the predictions of DW theory~\citep[e.g.,][]{efr11}.
The multi-arm and flocculent spirals do not fall under this context.
To ensure that a spiral galaxy is of the grand design type,
its mass structure has to be analyzed~\citep{mar13}.\footnote{In the early stages of this research~\citep{MarPue14},
we obtained two-dimensional stellar mass maps~\citep{zib09} by comparing photometry
with a Monte Carlo library of stellar population synthesis models, and dust radiation
models for longer wavelengths~\citep{daC08}. We obtained a filamentary mass structure in the spiral arms
of some objects, e.g., NGC~5194.
The structure is coincident with dust lane locations in the optical,
and is preserved even if only optical colors are adopted in the method.
A possible explanation may come from mass and dust relations,
derived from studies of nearby spiral galaxies~\citep{bri04,gro13,zah13}, which show a similar behavior.
Also in the case of NGC~5194, \citet{men12} suggest that stellar and dust components may be coupled.
More research will be conducted in the future in this respect.}
This is due to the fact that even in the NIR wavelengths the old population
can be significantly contaminated with young stellar objects, and dust
radiation for wavelengths longer than $2.5\micron$~\citep[e.g.,][]{mei12}.

To tackle this problem we take into account the fact that strong ``mass spirals''
must produce strong ``large-scale spiral arm shocks'' and the corresponding dust lanes~\citep{rob69,sly03,git04}.
Dust lanes can be adequately traced near spiral arms because of the extinction of
radiation that reveals them~\citep[see e.g.,][]{gon96}.

In order to trace the dust lanes we obtain maps of the $(g-3.6\micron)$\footnote{All the objects in our sample have
cross-sections in SDSS DR8~\citep{aih11}, expect for NGC~6946 for which we use the $V$-band~\citep{ken03},
and obtain the $(V-3.6\micron)$ color.} 
color for each object in our sample~(see e.g., figure~\ref{M51_dustlanes}).
For each of these images we apply the 2D (two-dimensional) Fourier transform method commonly
used to determine spiral arms pitch angle~\citep[e.g.,][]{con88,pue92,sar94,dav12,sav12}.
The Fourier amplitudes are given by

\begin{equation}
A(m,p)=\int_{u_{min}}^{u_{max}} \int_{-\pi}^{\pi}
I(u,\theta)e^{-i(m\theta+pu)}d\theta du,
\end{equation}

\noindent where $u=\ln R$; $R$ and $\theta$ are the polar coordinates;
$I(u,\theta)$ is the intensity at coordinates $\ln{R}$, $\theta$; $m$ is the
number of spiral arms (or modes); and $p$ is related to the spiral
arms pitch angle ($P$) by

\begin{equation}
 \tan{P} = -m/p_{\mathrm{max}},
\end{equation}

\noindent where $p_{\mathrm{max}}$ corresponds to the maximum of $A(m,p)$
where $m=0,1,2,3\dots$, i.e. the maximum of the Fourier spectrum for mode $m$.
Therefore, once the radial annulus to be analyzed is chosen by fixing $u_{min}$ and $u_{max}$,
the amplitude of the complex matrix $A(m,p)$ will show the more probable
pitch angle $P$ of that $m$ structure in that annulus.

The Fourier amplitudes for modes $m=1$ through $m=6$ were determined as a function of the
frequency $p$. The adopted radial ranges, wherein the spiral structure is contained,
are shown in table~\ref{tbl-2}.
The method gives negative or positive pitch angles ($P$) depending on the
``S'' or ``Z'' on-the-sky view of the object.\footnote{
It should be noted that all $p$ frequencies, positive and negative,
can have an important role when spiral arm modulation is studied~\citep{pue00}.}

The purpose of applying the 2D Fourier transform method was to identify
which objects host strong mass spirals with $m=2$. To achieve this goal we
utilize the ``footprints'' that the gravitational potential produces in the gas
as it shocks. In figure~\ref{FFT_gIRAC1} we show the results of this analysis
for the $(g-3.6\micron)$ images. According to this result, only
three of the objects, NGC~628, NGC~3627, and NGC~5194, present
evidence to host strong mass spirals with $m=2$ in their disks.
Less clear evidence can also be appreciated for NGC~3351 and NGC~5055.
In the case of NGC~5194 the mode $m=4$ is competing in amplitude
with the $m=2$ mode, although with a smaller pitch angle.
For comparison we apply the same method to the $3.6\micron$ image and plot
the results in figure~\ref{fFFT_IRAC1only}. Significant differences
can be appreciated for some objects, e.g., NGC~2841, and NGC~6946.
As explained before, the radiation of the old stellar population
as seen in the $3.6\micron$ image is accompanied by radiation of young stellar objects,
and dust emission~\citep{mei12}. In this sense, the mass distribution that can be inferred from the $3.6\micron$ image
would not be the same to that required to produce the dust lanes on the $(g-3.6\micron)$ image,
i.e., the product of the spiral shock. The difference between figures~\ref{FFT_gIRAC1} and~\ref{fFFT_IRAC1only}
is more probably due to the fact that the NIR images do not trace solely the stellar mass.

This preliminary test gives us information of which objects
present mass spirals with strong $m=2$ modes in their disks,
and, if this is the case, are likely to have a constant spiral pattern speed for all radii.
For the rest of the objects, we anticipate that a lack of positive results,
regarding offsets for the different stages of the star formation sequence,
is probably due to the absence of adequate spiral perturbations causing the expected phenomenon.


\begin{deluxetable}{lccr}
\tabletypesize{\scriptsize}
\tablecaption{Additional parameters~\label{tbl-2}}
\tablewidth{0pt}
\tablehead{
\colhead{Object} & 
\colhead{$R_{\rm{end}}(\arcsec)$} &
\colhead{$R_{\rm{end}}(\rm{kpc})$} &
\colhead{$\Delta{R}(\arcsec)$}
}

\startdata


NGC~628  	&	190	&	9.1  	&	(26.3-223.0)	\\
NGC~2403	&	368	&	8.1   	&	(28.9-348.3)	\\
NGC~2841	&	158	&	9.4	&	(31.5-199.0)	\\
NGC~3351	&	158	&	6.9	&	(95.0-214.7)	\\
NGC~3521	&	115	&	4.5	&	(35.8-125.0)	\\
NGC~3627	&	165	&	5.2  	&	(63.3-189.4)	\\
NGC~5055	&	180	&	7.3   	&	(52.4-288.4)	\\
NGC~5194	&	270	&	11.9    &	(43.5-295.7)	\\
NGC~6946	&	195	&	5.2	&	(57.4-243.9)	\\

\enddata

\tablecomments{
Columns 2 and 3: maximum radial extent of the spiral arms, in arcseconds and kpc respectively,
determined visually from the $3.6\micron$ image~\citep{ken03}.
Column 4: adopted radial ranges ($R_{min}-R_{max}$) in arcseconds, for the 2D Fourier transform analysis
(see section~\ref{mass_spirals}).
}

\end{deluxetable}


\section{Results and discussion}~\label{results}

We compare the phases obtained for the 9 objects in our sample with the expected
star formation sequence. We find that only two of the objects,
NGC~628 \& NGC~5194, present coherent phases similar to the predictions of our
model (equation~\ref{eqTHETA}, and figure~\ref{fig_shifts}).
For the rest of the objects the plots show dissimilar patterns as compared
to the model~(see also section~\ref{mass_spirals}). Figures~\ref{N628_HICO24UV} and~\ref{N5194_HICO24UV}, show the phases plots
for NGC~628 and NGC~5194 respectively, as a function of radius. The mean radius, $R_{\rm{mean}}$,
has been normalized to the radius where the spirals are observed to end in the $3.6\micron$ image~\citep[SINGS,][]{ken03}.
These ``spiral end points'' ($R_{\rm{end}}$) are listed in table~\ref{tbl-2}.
The long-dashed lines, in the right panels of the figures, indicate the $\pm~1-\sigma$ error as a function of radius, this takes into account
the resolution of the data (FWHM$=13\arcsec$), and the median error from the Montecarlo simulations (see section~\ref{error_analysis}).

For the NGC~628 H{\rm{I}} data, we did not obtained Fourier phases that resemble our theoretical
model (i.e., equation~\ref{eqTHETA}, and figure~\ref{fig_shifts}).
The CO, $24\micron$, and FUV phases seem to have the expected order of the star formation sequence for
the inner radii (before an hypothetical corotation). However, only the CO and $24\micron$ phases show
an indication of a corotation around 
$0.46<R_{\rm{mean}}/R_{\rm{end}}>0.48$ ($87\arcsec<R>91\arcsec$).
Before this region, most of the CO$-24\micron$ phase shifts show positive values as expected. The agreement with expectations
is not fulfilled after $R_{\rm{mean}}/R_{\rm{end}}\sim0.6$ ($R\sim114\arcsec$), where positive CO$-24\micron$ phase shifts
are obtained instead of negative ones.
The phase shifts indicate that NGC~628's strong spirals may terminate near the
suggested corotation~\citep[see also][]{con86,pat91},
if we assume a constant spiral pattern speed of the DW type.
We consider the evidence of having $\Omega_{p}\sim$~constant for NGC~628 as moderate~\citep[see also][]{ced13}.

For NGC~5194 we again did not obtain phases values coincident with expectations for the H{\rm{I}} data.
Once more the CO, $24\micron$, and FUV phases present an ordered pattern that resembles
the anticipated model (equation~\ref{eqTHETA}, and figure~\ref{fig_shifts}).
The CO$-24\micron$ phase shifts suggest a corotation zone near $0.74<R_{\rm{mean}}/R_{\rm{end}}>0.76$
($200\arcsec<R>205\arcsec$).
Before this zone most of the CO$-24\micron$ phase shifts show positive values, and negative values
are obtained afterwards.
The CO$-$FUV phase shifts also feature a corotation radius near $R_{\rm{mean}}/R_{\rm{end}}\sim0.63$ ($R\sim170\arcsec$).
In figure~\ref{offsets_M51} we show that the size of the CO$-24\micron$ angular offsets are of the same
magnitude as those obtained by~\citet{sch13}, who adopt a polar cross-correlation method.
\citet{sch13} take advantage of the PdBI Arcsecond Whirlpool Survey (PAWS), mapping $^{12}$CO~$J=\jone$.
We consider a clear evidence of having $\Omega_{p}\sim$~constant for NGC~5194.

The case of the barred galaxy NGC~3627 is shown in figure~\ref{N3627_HICO24UV}.
It has been shown that this object hosts strong ($m=2$) mass spirals in its disk (see section~\ref{mass_spirals}).
For the CO-FUV phase shifts there is a hint of a corotation around $R_{\rm{mean}}/R_{\rm{end}}\sim0.45$ ($R\sim74\arcsec$).
However the phase shifts for CO$-$H{\rm{I}, and CO$-24\micron$ do not confirm this possibility.
The bar length is $\sim49\arcsec$, or $R_{\rm{mean}}/R_{\rm{end}}\sim0.3$
for $R_{\rm{end}}\sim165\arcsec$.

As a supplementary study we also analyzed the H$\alpha$ and $3.6\micron$ data from SINGS~\citep{ken03},
for NGC~628 and NGC~5194, as well as the $70\micron$, $160\micron$, and $250\micron$ data for NGC~5194,
the latter obtained from the VNGS (The Very Nearby Galaxy Survey\footnote{http://hedam.lam.fr})
that has used the PACS and SPIRE instruments on Herschel Space Observatory~\citep{pil10}.
We compare the H$\alpha$ and $3.6\micron$ phases with the CO. Figures~\ref{N628_COHa3p6} and
\ref{N5194_COHa3p6} show these complementary phases plots for NGC~628 and NGC~5194 respectively.
The CO-H$\alpha$ phase shift plots show a very similar behavior as compared with the CO$-24\micron$ phase shifts.
In the case of NGC~628, the CO$-3.6\micron$ phase shifts show no indication of a corotation.
For NGC~5194 the CO$-3.6\micron$ phase shift may be consistent with expectations for
$R_{\rm{mean}}/R_{\rm{end}}>0.5$ ($R\sim135\arcsec$),
if we assume that the $3.6\micron$ comes solely from an old stellar population downstream the shock?
Albeit the local $3.6\micron$ radiation can be strongly affected by dust
emission and young stellar components~\citep{mei12}.

In the case of the analysis of NGC~5194's Herschel data, we convolved the HERACLES CO, PACS $70\micron$, and PACS $160\micron$
images with a Gaussian function having a FWHM of $18.1\arcsec$ (the resolution of the SPIRE $250\micron$ image).
The results are shown in figure~\ref{N5194_COHerschel}.
We will discuss the region within $0.3<R_{\rm{mean}}/R_{\rm{end}}>0.9$~($81\arcsec<R>243\arcsec$).
Before this region, the phase shifts are within the data resolution,
and to larger radii we approach the zone of influence of NGC 5195.
The phase shifts between CO$-250\micron$, CO$-160\micron$, and CO$-70\micron$, show
a transition zone from being mostly positive to negative
around $R_{\rm{mean}}/R_{\rm{end}}\sim0.7$ ($R\sim189\arcsec$).
This is another indication of a possible corotation near this zone.
The figure also indicates that, for most of the spiral arm,
the sequence of these tracers is arranged so that the
CO is prior the $160\micron$ emission, and followed by the $70\micron$ emission.
For the $250\micron$ emission the situation is unclear, it seems
to be located prior the $160\micron$ emission before the corotation region,
and later the $70\micron$ after the corotation region.


\subsection{Error analysis}~\label{error_analysis}

In this section we analyze the sources of uncertainty that
arise by using the Fourier phases method to analyze spiral perturbations.
There are mainly three sources of error that may introduce biases
in the results. These are the $\pi$ radians symmetry assumption (i.e., $m=2$),
the inherent uncertainties in the intensities, and the projection parameters.

To analyze the errors that the $\pi$ radians symmetry assumption ($m=2$)
has on our results, we isolate the spiral arms
of NGC~628 and NGC~5194 according to figures~\ref{M74_arms}, and~\ref{M51_arms}
respectively. Then, the isolated spiral arms were treated as $m=1$ modes
and analyzed with the Fourier technique.
For NGC~628, the results of this test are shown on figure~\ref{N628_isolarms}.
For the spiral arm listed as number ``1'', the phase shifts exhibit a similar behavior
to the $m=2$ mode analysis (figure~\ref{N628_HICO24UV}). Albeit the transition
from positive CO$-24\micron$ phase shifts to the corotation, $R_{\rm{mean}}/R_{\rm{end}}\sim0.47$ ($R\sim89\arcsec$),
bears a better resemblance, and less scatter, as compared to the expected model (equation~\ref{eqTHETA}, and figure~\ref{fig_shifts}).
For the spiral arm listed as number ``2'', the phase shifts behavior
is more chaotic in comparison with expectations.
The results for NGC~5194 are shown in figure~\ref{N5194_isolarms}.
For the spiral arm ``1'' the phase shifts show no clear
corotation region. Contrariwise, the CO$-24\micron$ phase shifts for spiral arm ``2''
present a very good resemblance to the expected model.
From these tests we can conclude that, in the case for NGC~628 and NGC~5194,
the two arms presents dissimilar characteristics.
Nonetheless, by adopting a $\pi$ radians symmetry ($m=2$) we are basically obtaining
an ``average'' result for both arms, which still indeed shows an indication of a corotation.

The errors related to the inherent uncertainties in the intensities
can be analyzed with Monte Carlo methods. For this purpose we adopt the
HERACLES~\citep[CO,][]{ler09} uncertainty maps. 
Each pixel of the integrated intensity maps was modeled with a Gaussian
distribution with a standard deviation equal to its respective uncertainty.
We generate 100 images by randomly assigning an intensity value
to each pixel in the CO map in accordance to the Gaussian model.
For each of the 100 frames we compute the Fourier phases values (see section~\ref{analysis}}).
Then we obtained a probability distribution of the phases
for each analyzed radius (see figure~\ref{phases_prob_M51}).
From the probability distributions we estimate the $1-\sigma$
error for each radius. The outcomes are shown in figure~\ref{one_sigma_err}
for NGC~628, and NGC~5194. From these radial distributions we obtain
the median $1-\sigma$ error. This median error was added in quadrature to the
error corresponding to the data resolution. 
We assume that the data from the other analyzed wavebands have similar uncertainties.

To asses the errors that the projection parameters may introduce in our analysis
we modified the inclination angles ($\alpha$) by $\pm5\degr$, and the position angles (PA) by $\pm10\degr$.
The phases were recalculated. We show the results for NGC~5194 in figure~\ref{proyec_param}.
It was found that by varying the projection parameters we introduce small differences
in the values of the phases, however, the overall trends and spatial distributions
are much the same. This means that the ``real'' values for the phases are certainly
affected by the ``correct'' projection parameters, but these uncertainties
do not affect the main results of our analysis. The same conclusions are achieved for NGC~628.

\subsection{UV emission}

The UV radiation from young stellar objects belongs mainly to stars
with spectral type O. These stars are short-lived ($\sim1-3$ Myr),
hence they are never found too far away from their birthplaces, which can be strongly affected
by dust attenuation. By the time when dust has been photo-evaporated locally,
O stars may cease to exist and their unobscured UV radiation would be significantly reduced for that region.
A possible example of this phenomenon may be that of NGC~5195 (the companion of NGC~5194).
NGC~5195 is not detected in the FUV~\citep{kuch00} but only in the NUV.
In the case of our research, we would not expect to find important UV radiation as expected in the ordered
sequence of figure~\ref{fig_shifts}. This may be occurring for the
results obtained for NGC~628 and NGC~5194, where the FUV Fourier phases partially
behave as expected. O stars radiation may also heat the molecules of CO (as they do with the dust) affecting
its emission (Rosa Gonz\'alez-L\'opezlira, private communication 2013).
This may explain the ``bumps'' in the CO$-24\micron$ phase shifts obtained in
figures~\ref{N628_HICO24UV}, and~\ref{N5194_HICO24UV}, for NGC~628, and NGC~5194 respectively.

\subsection{Spiral pattern speed dependence with radius}

Our result for a constant spiral pattern speed for NGC~5194 differs from
the findings of~\citet{mei08}, who applied the radial Tremaine-Weinberg (TWR) method
and found a dependence of $\Omega_{p}$ with radius~\citep[see also][]{mar09b}.
The validity of the TWR method is based on the use of a kinematic tracer (gas)
that must obey the continuity equation~{\it{in the plane}} of the galaxy.
However, the gas kinematics in the spiral arms
and interarms of NGC~5194, indicate that out-of-plane motions may be significant~\citep{she07}.
Moreover, the spiral shock can extend to heights above the galactic midplane~\citep{mar99,alf01}.
The velocity vector towards the observer would be affected by these circumstances, 
and in the same way the continuity equation in the plane of the galaxy.

\section{Conclusions}

The results for the H{\rm{I} distribution near the spiral arms of NGC~628 and NGC~5194,
indicate, once again, that H{\rm{I} is a photodissociated product of H$_2$ molecules
near star forming regions~\citep{all86,til88,ran92,lou13}. In a DW scenario
it would not entirely trace the highly compressed gas near the dust lanes.
Thus its location in figure~\ref{fig_shifts} is not entirely accurate.

The results obtained for NGC~5194 in this research are consistent with
the findings of, e.g.,~\citet{vog88},~\citet{tos02}, and~\citet{pat06},
where the CO emission is preceding the star formation across the spiral arms.
\citet{egu09}~already reported dissimilar CO-H$\alpha$ offsets for the two different
spiral arms. This may be due to the gravitational influence of the companion.
In the case of NGC~628,~\citet{gus13} and~\citet{gus14},
studied the different behavior of the star formation
for the two distinct spiral arms. They indicate that this feature
is seen for other grand design spirals, and the reason for this is not fully
understood. Why is the star formation dissimilar for the two
distinct arms in some grand-design spirals? 
The answer to this question may have to do with the origin of the
(grand design) spiral structure which is not fully understood yet.

In summary our results indicate that two of the nine analyzed objects
present a constant spiral pattern speed with a main corotation,
as inferred from the comparison of observations and the
star forming sequence model (equation~\ref{eqTHETA}
and figure~\ref{fig_shifts}, excluding the H{\rm{I} data), across spiral arms.
These are NGC~628 with a corotation $R_{CR}\sim89\pm2\arcsec$ (see figure~\ref{M74_arms}),
and NGC~5194 with a corotation $R_{CR}\sim202\pm3\arcsec$ (see figure~\ref{M51_arms}).
By assuming a flat rotation curve, and adopting the parameters in table~\ref{tbl-1}, we have spiral pattern
speeds of $\Omega_{p}\sim41.8\pm1$ (km s$^{-1}$ kpc$^{-1}$)
for NGC~628, and $\Omega_{p}\sim23.7\pm0.5$ (km s$^{-1}$ kpc$^{-1}$) for NGC~5194.
The evidence of having $\Omega_{p}\sim$~constant is more clear for NGC~5194 than for NGC~628.
The only object with strong two-arm ($m=2$) spiral structure not
showing signs of a star formation sequence across its spiral arms is NGC~3627.
For the remaining galaxies it was found that they do not
present coherent phases, a fact that can be explained by the absence
of a strong two-arm ($m=2$) mass structure in their spirals.

\acknowledgments

We appreciate discussions with Rosa Gonz\'alez-L\'opezlira.
We acknowledge the anonymous referee for his/her constructive remarks.
EMG acknowledges CONACYT's financial support for postdoctoral fellowship at INAOE.
IP acknowledges support from the Mexican foundation CONACYT.

This work made use of SINGS, `The Spitzer Infrared Nearby Galaxies Survey'~\citep{ken03},
the Galaxy Evolution Explorer (GALEX) `Ultraviolet Atlas of Nearby Galaxies'~\citep{gil07},
`The H{\rm{I} Nearby Galaxy Survey'~\citep[THINGS,][]{wal08}, and
the `HERA CO-Line Extragalactic Survey'~\citep[HERACLES,][]{ler09}.
The VNGS data, `The Very Nearby Galaxy Survey', was accessed through the Herschel Database in Marseille
(HeDaM - http://hedam.lam.fr) operated by CeSAM and hosted by the Laboratoire d'Astrophysique de Marseille.
We acknowledge the usage of the HyperLeda database (http://leda.univ-lyon1.fr).
This research has made use of the NASA/IPAC Extragalactic Database (NED) which is operated by the Jet Propulsion Laboratory,
California Institute of Technology, under contract with the National Aeronautics and Space Administration.


\begin{figure*}
\centering
\includegraphics[scale=0.6,angle=0]{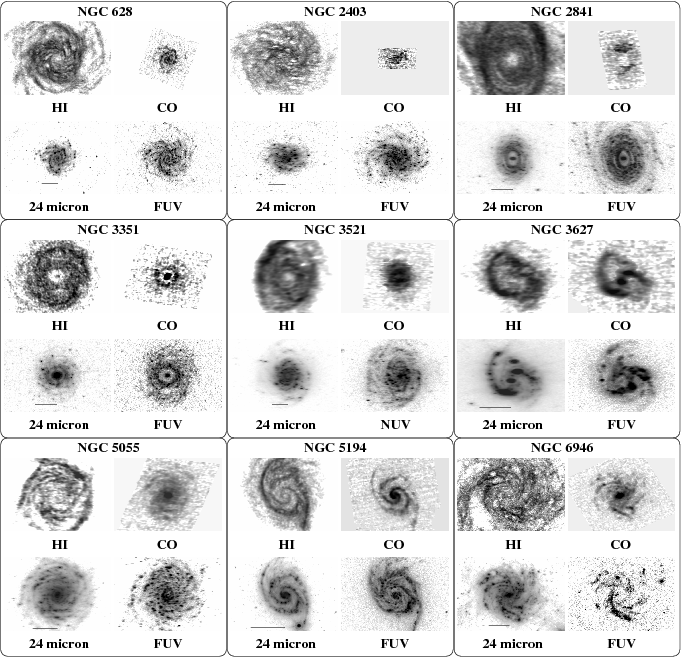}
\caption[f1]{Deprojected images of
H{\rm{I}} from THINGS~\citep{wal08},
CO from HERACLES~\citep{ler09},
$24\micron$ from SINGS~\citep{ken03},
and FUV from GALEX~\citep{gil07}.
For each object the scale is the same in the four images.
Display is in logarithmic scale.
Data are shown for NGC~628, NGC~2403, NGC~2841,
NGC~3351, NGC~3521, NGC~3627, NGC~5055, NGC~5194, and NGC~6946.
The horizontal line, in the $24\micron$ frames, represents the maximum radial extent of the spiral arms,
$R_{\rm{end}}$, measured from the galaxy's center (cf. table~\ref{tbl-2}).
~\label{fig_sample}}
\end{figure*}

\begin{figure*}
\centering
\includegraphics[scale=0.8,angle=0]{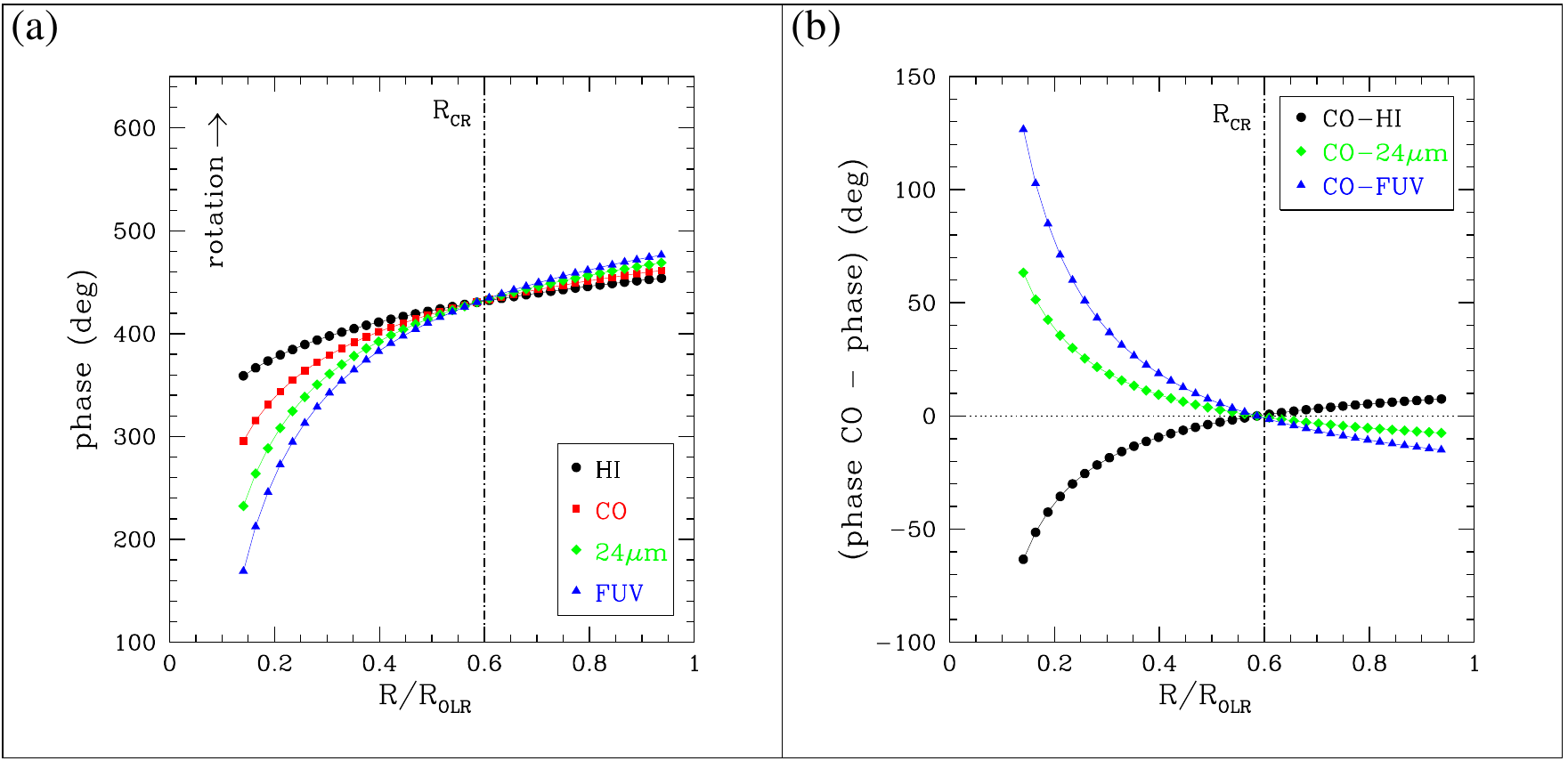}
\caption[f2]{{\it~Panel (a):} toy model illustration showing the expected azimuthal phases in degrees ($y$-axis) for the H{\rm{I}} (circles),
CO (squares), $24\micron$ (diamonds), and FUV data (triangles). In this hypothetical scenario star formation is shock-induced
by a spiral pattern, of the trailing type, with a constant angular velocity for all radii.
The vertical dotted-dashed line indicates the corotation radius, $R_{CR}$, where the phases invert their order. 
{\it~Panel (b):} illustration showing the expected phase shifts ($y$-axis), CO$-$H{\rm{I}} (circles),
CO$-24\micron$ (diamonds), and CO$-$FUV (triangles). In both panels, the radius $R$ ($x$-axis),
is normalized to the Outer Lindblad Resonance (OLR) radius, $R_{OLR} = R_{CR} (1+\frac{\sqrt{2}}{2}) $,
assuming a flat rotation curve.
~\label{fig_shifts}}
\end{figure*}

\begin{figure*}
\centering
\includegraphics[scale=0.6,angle=0]{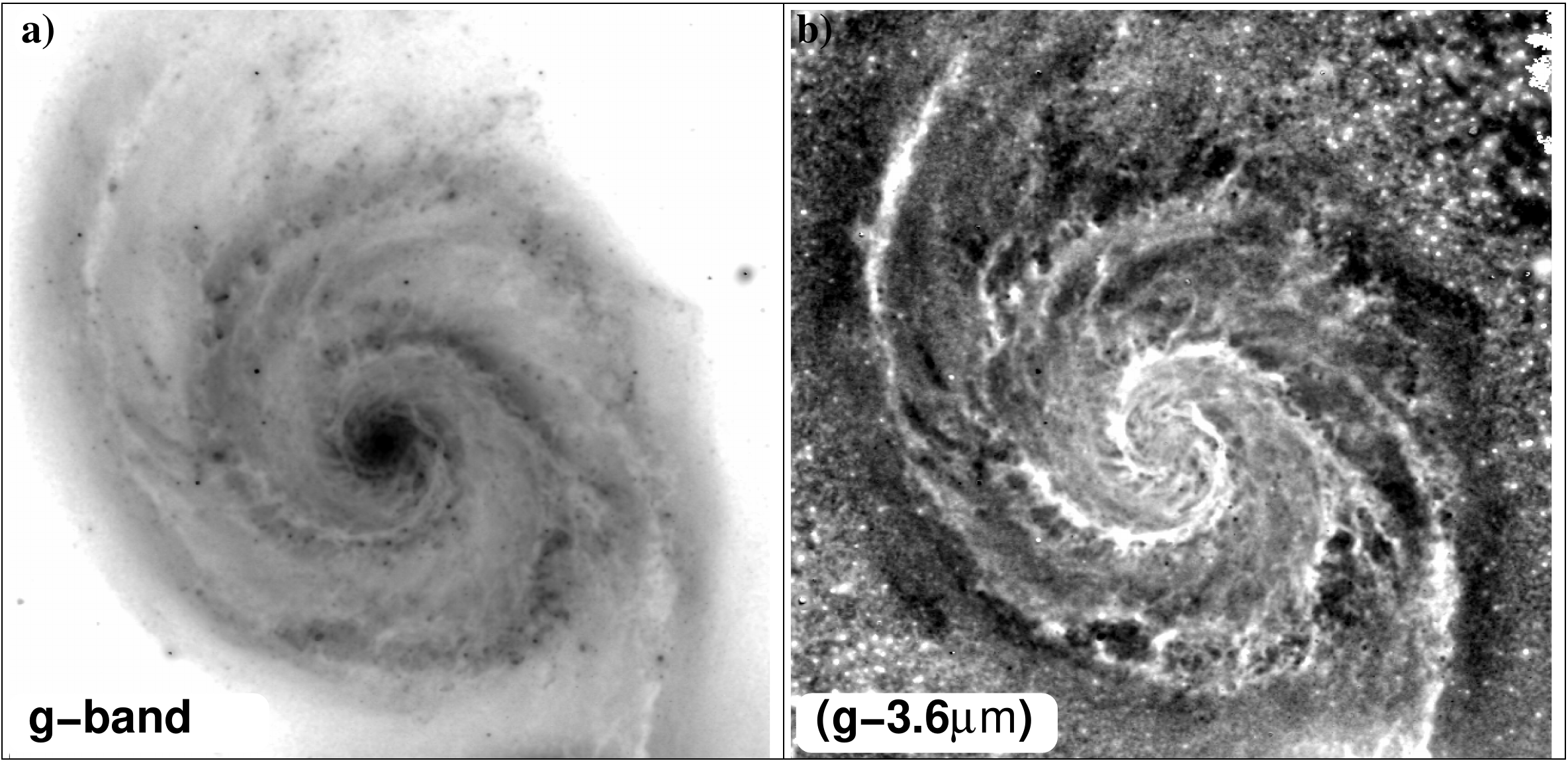}
\caption[f3]{{\it~Panel (a):} NGC~5194 deprojected $g$-band (SDSS). Dust lanes
can be appreciated as extinction features (with less brightness for this color display)
in the concave side of the arms.
{\it~Panel (b):} The $(g-3.6\micron)$ color map for NGC~5194. Dust lanes are traced
as less intense features (with the adopted color display).
An almost identical resemblance to the dust lanes of panel (a) can be appreciated.
~\label{M51_dustlanes}}
\end{figure*}

\begin{figure*}
\centering
\includegraphics[scale=0.8,angle=0]{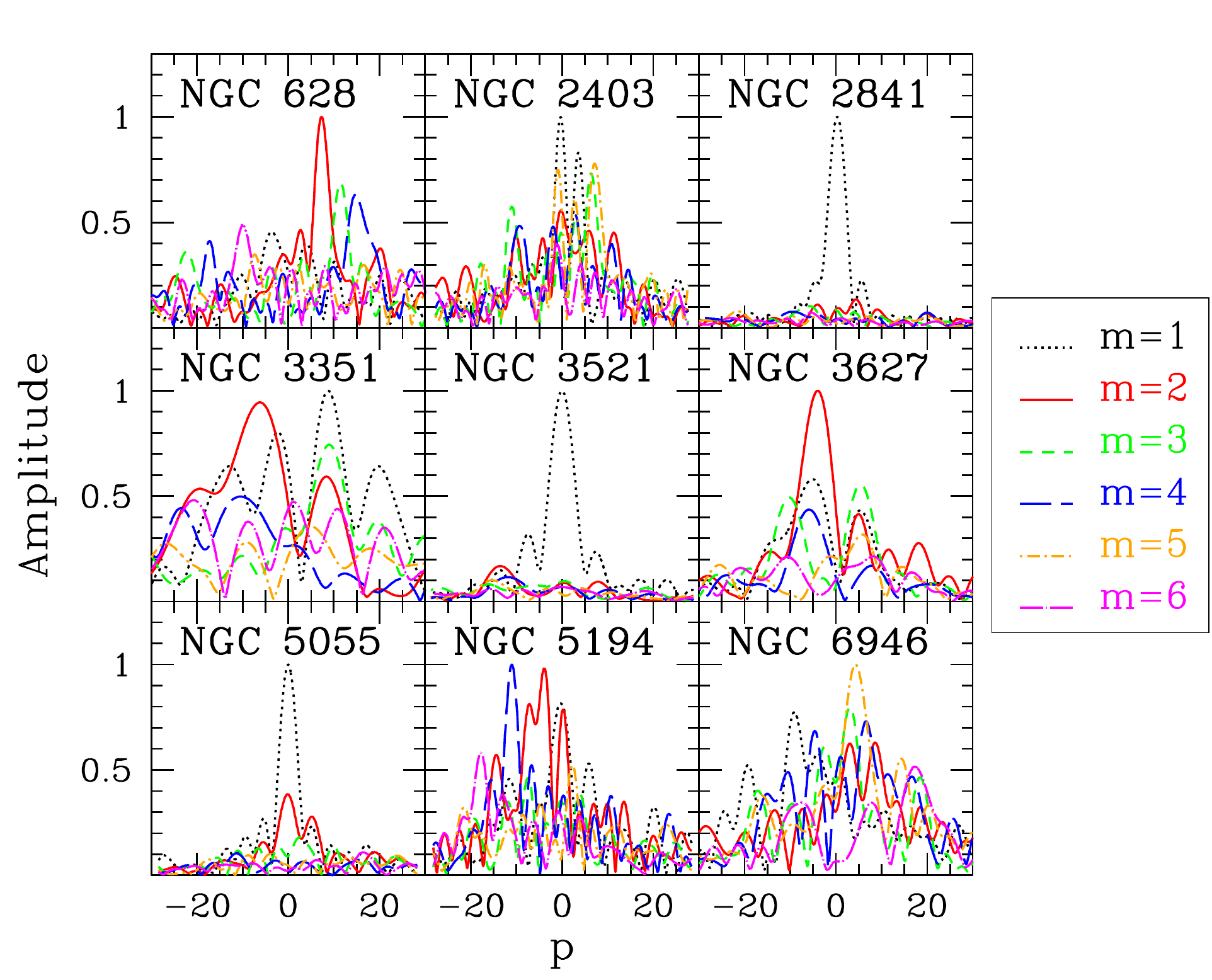}
\caption[f4]{Amplitudes vs. $p$ frequencies (related to the pitch angle),
obtained by applying the 2D Fourier transform method to the ($g-3.6\mu\mathrm{m}$) images.
Modes $m=1$ through $m=6$ are plotted.
~\label{FFT_gIRAC1}}
\end{figure*}

\begin{figure*}
\centering
\includegraphics[scale=0.8,angle=0]{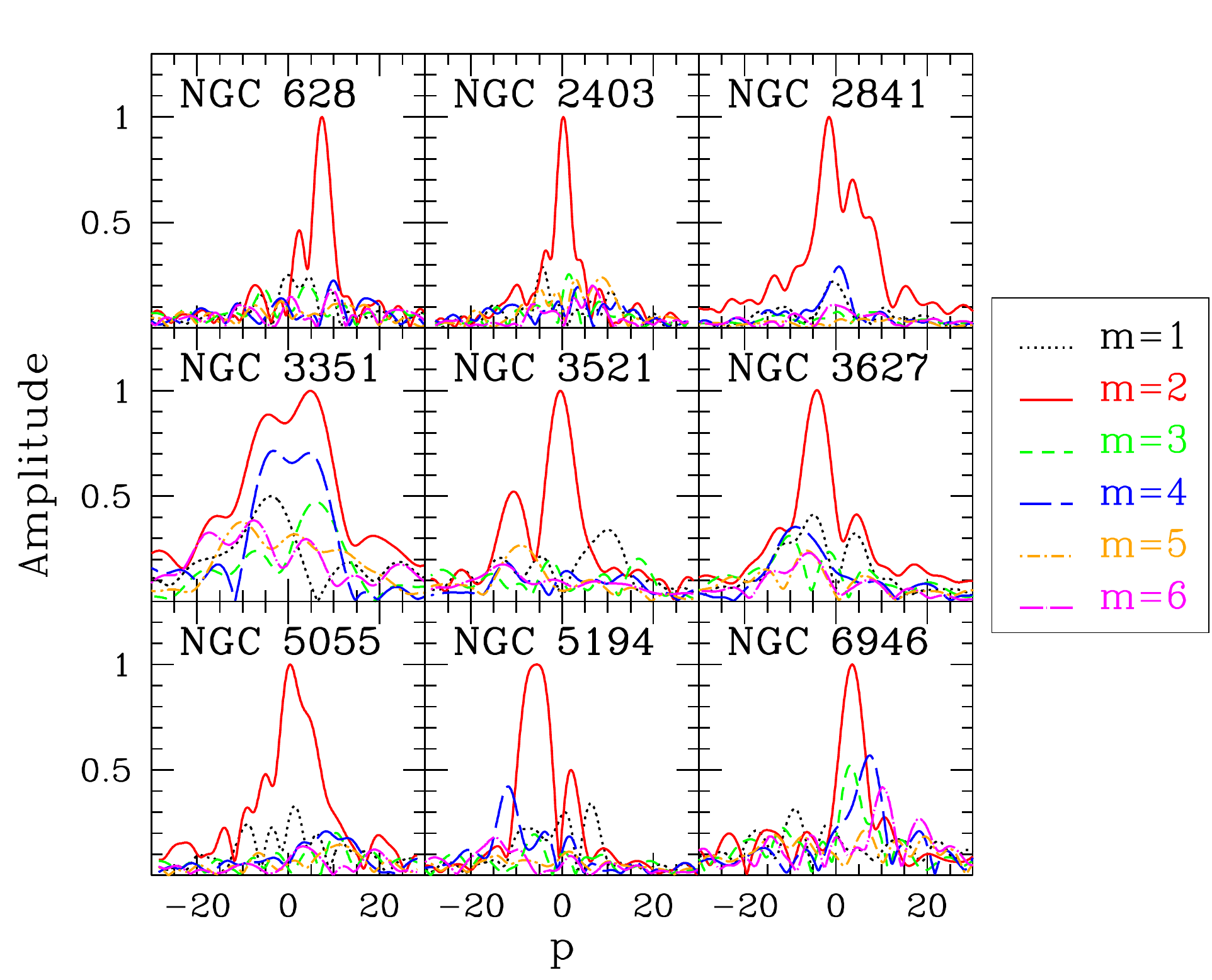}
\caption[f5]{Amplitudes vs. $p$ frequencies, obtained by applying the 2D Fourier method to the $3.6\mu\mathrm{m}$
(IRAC channel 1) images. Modes $m=1$ through $m=6$ are plotted.
~\label{fFFT_IRAC1only}}
\end{figure*}

\begin{figure*}
\centering
\includegraphics[scale=0.8,angle=0]{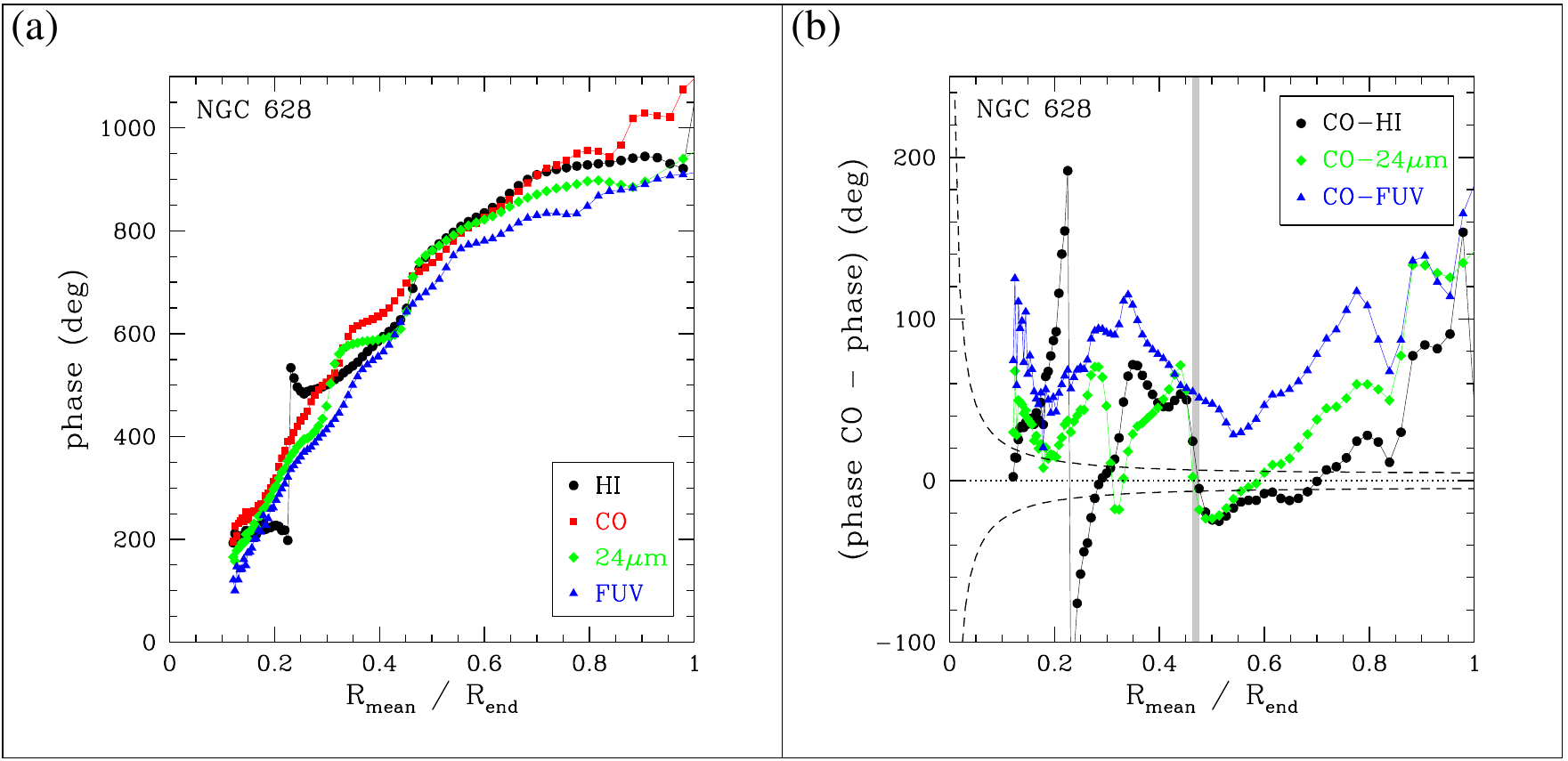}
\caption[f6]{Results for NGC~628 (H{\rm{I}}, CO, $24\micron$, \& FUV). {\it~Panel (a):} azimuthal phases in degrees ($y$-axis), vs.,
normalized radius $R_{\rm{mean}}/R_{\rm{end}}$ ($x$-axis).
Circle figures represent the H{\rm{I}}, square figures the CO, diamond figures the $24\micron$, and triangle figures the FUV emission.
{\it~Panel (b):} phase shifts between the phases in CO and other tracers ($y$-axis),
vs., normalized radius.
Circle figures represent the phase shift between CO and H{\rm{I}}, diamonds figures the phase shift between CO and $24\micron$,
and triangle figures the phase shift between CO and FUV.
The gray shaded region shows the proposed corotation, $R_{\rm{mean}}/R_{\rm{end}}\sim0.47$, for the CO-$24\micron$ phase shifts.
The long-dashed line indicates the $\pm1-\sigma$ error.
~\label{N628_HICO24UV}}
\end{figure*}

\begin{figure*}
\centering
\includegraphics[scale=0.8,angle=0]{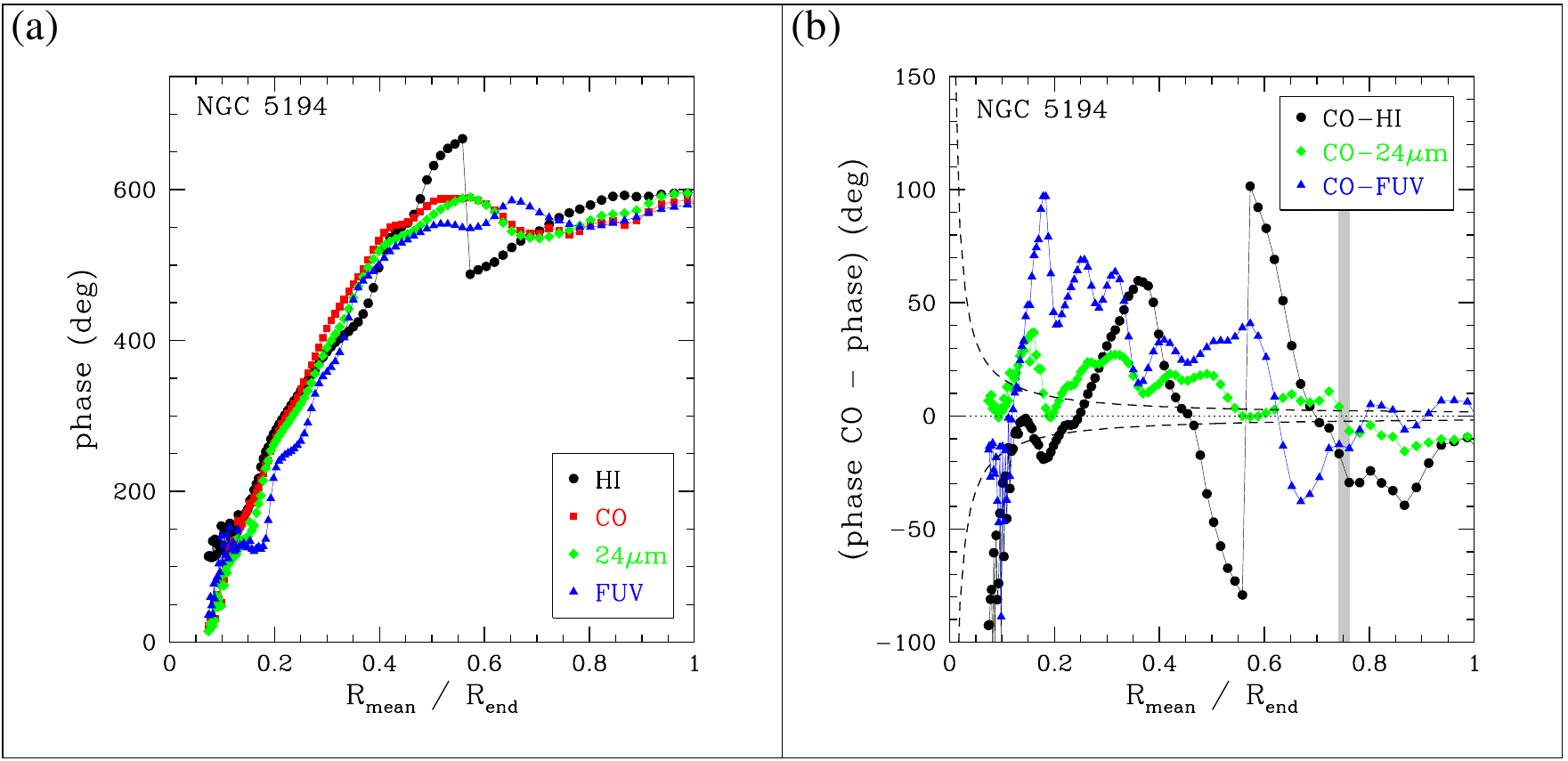}
\caption[f7]{Results for NGC~5194 (H{\rm{I}}, CO, $24\micron$, \& FUV).
{\it~Panel (a):} azimuthal phases in degrees ($y$-axis), vs.,
normalized radius $R_{\rm{mean}}/R_{\rm{end}}$ ($x$-axis).
{\it~Panel (b):} phase shifts between CO and other tracers ($y$-axis),
vs., normalized radius.
The gray shaded region shows the proposed corotation, $R_{\rm{mean}}/R_{\rm{end}}\sim0.75$, for the CO-$24\micron$ phase shifts.
Same symbolism as in figure~\ref{N628_HICO24UV}.
~\label{N5194_HICO24UV}}
\end{figure*}

\begin{figure*}
\centering
\includegraphics[scale=0.6,angle=0]{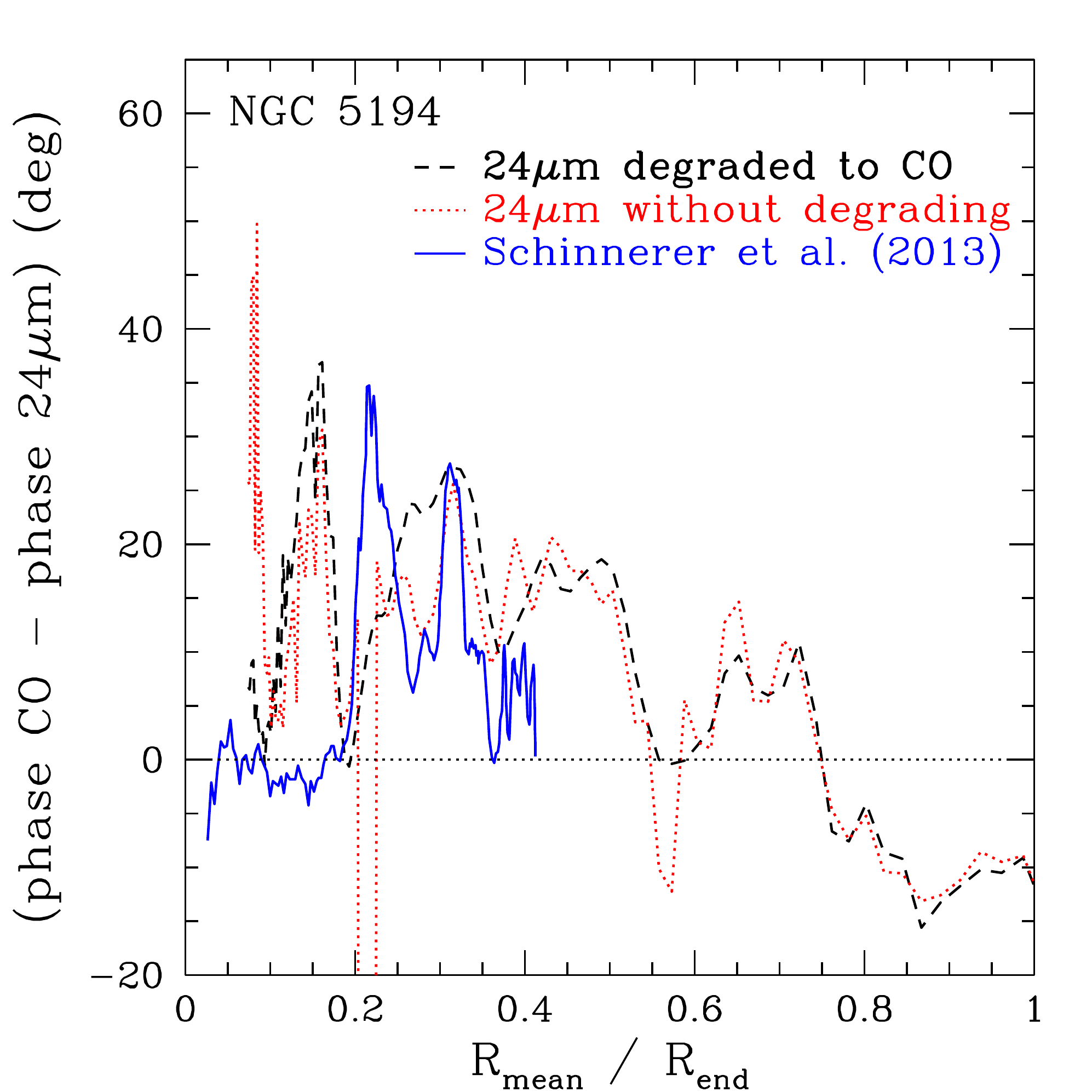}
\caption[f8]{Azimuthal offsets of CO$-24\micron$ for NGC~5194. The dashed line indicates
the offset obtained with the Fourier method described in section~\ref{analysis}. The dotted
line indicates the offsets obtained if the $24\micron$ data is not degraded to the CO resolution (13\arcsec).
The continuous line shows the offsets obtained by~\citet{sch13}. The magnitude of the
offsets obtained for the three lines is similar.
~\label{offsets_M51}}
\end{figure*}

\begin{figure*}
\centering
\includegraphics[scale=0.8,angle=0]{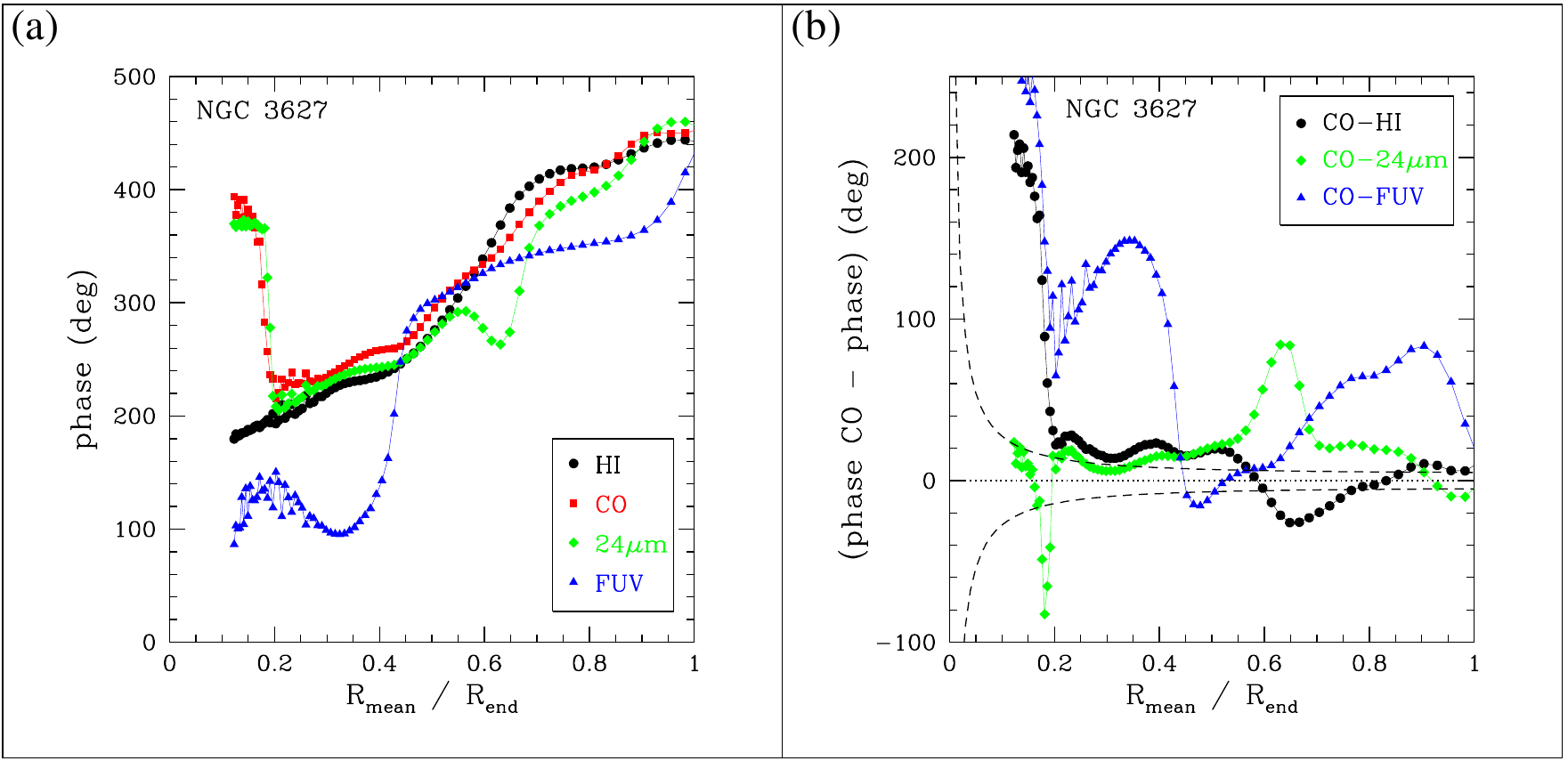}
\caption[f9]{Results for NGC~3627 (H{\rm{I}}, CO, $24\micron$, \& FUV).
{\it~Panel (a):} azimuthal phases in degrees ($y$-axis), vs.,
normalized radius $R_{\rm{mean}}/R_{\rm{end}}$ ($x$-axis).
{\it~Panel (b):} phase shift between CO and other tracers ($y$-axis),
vs., normalized radius.
Same symbolism as in figure~\ref{N628_HICO24UV}.
~\label{N3627_HICO24UV}}
\end{figure*}

\begin{figure*}
\centering
\includegraphics[scale=0.8,angle=0]{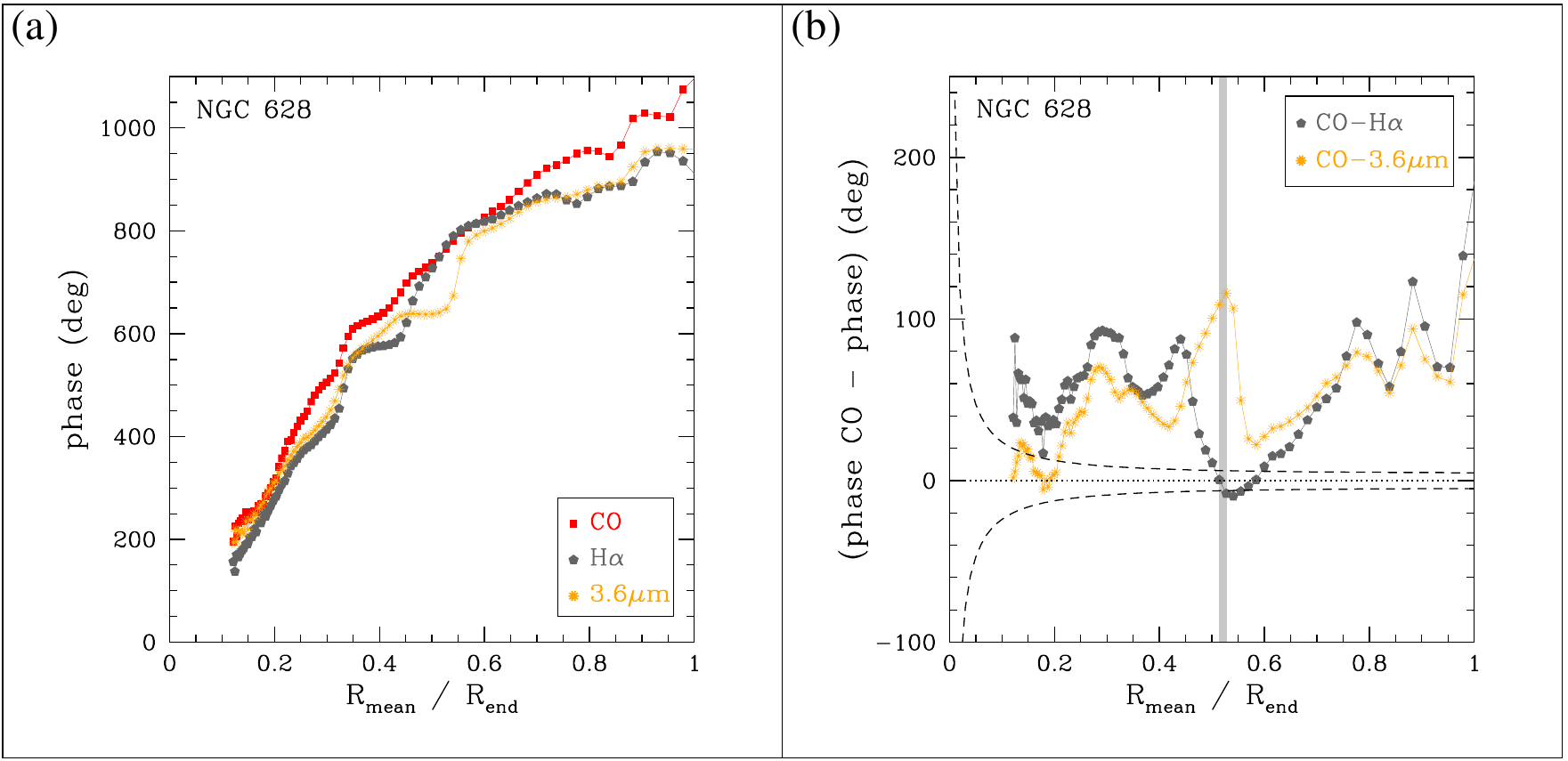}
\caption[f10]{Results for NGC~628 (CO, H$\alpha$, \& $3.6\micron$).
{\it~Panel (a):} azimuthal phases in degrees ($y$-axis), vs.,
normalized radius $R_{\rm{mean}}/R_{\rm{end}}$ ($x$-axis).
Square figures represent the CO, pentagon figures the H$\alpha$, and asterisk figures the $3.6\micron$ emission.
{\it~Panel (b):} phase shifts between CO and other tracers ($y$-axis),
vs., normalized radius.
Pentagon figures represent the phase shift between CO and H$\alpha$, and asterisk figures the phase shift between CO and $3.6\micron$.
The gray shaded region shows the proposed corotation, $R_{\rm{mean}}/R_{\rm{end}}\sim0.52$, for the CO-H$\alpha$ phase shifts.
The long-dashed line indicates the 1-$\sigma$ error.
~\label{N628_COHa3p6}}
\end{figure*}

\begin{figure*}
\centering
\includegraphics[scale=0.8,angle=0]{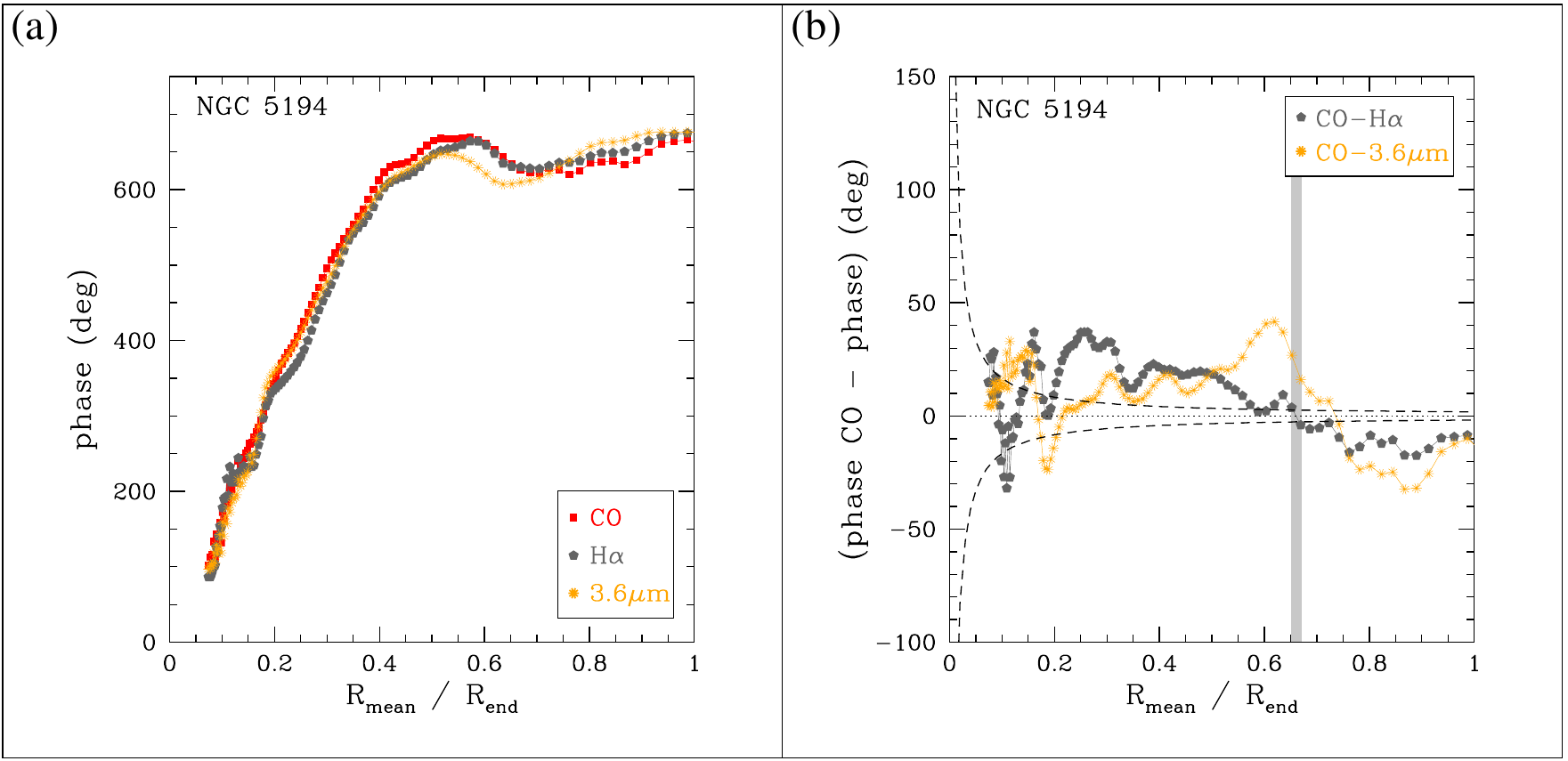}
\caption[f11]{Results for NGC~5194 (CO, H$\alpha$, \& $3.6\micron$).
{\it~Panel (a):} azimuthal phases in degrees ($y$-axis), vs.,
normalized radius $R_{\rm{mean}}/R_{\rm{end}}$ ($x$-axis).
{\it~Panel (b):} phase shift between CO and other tracers ($y$-axis),
vs., normalized radius.
The gray shaded region shows the proposed corotation, $R_{\rm{mean}}/R_{\rm{end}}\sim0.66$, for the CO-H$\alpha$ phase shifts.
Same symbolism as in figure~\ref{N628_COHa3p6}.
~\label{N5194_COHa3p6}}
\end{figure*}

\begin{figure*}
\centering
\includegraphics[scale=0.8,angle=0]{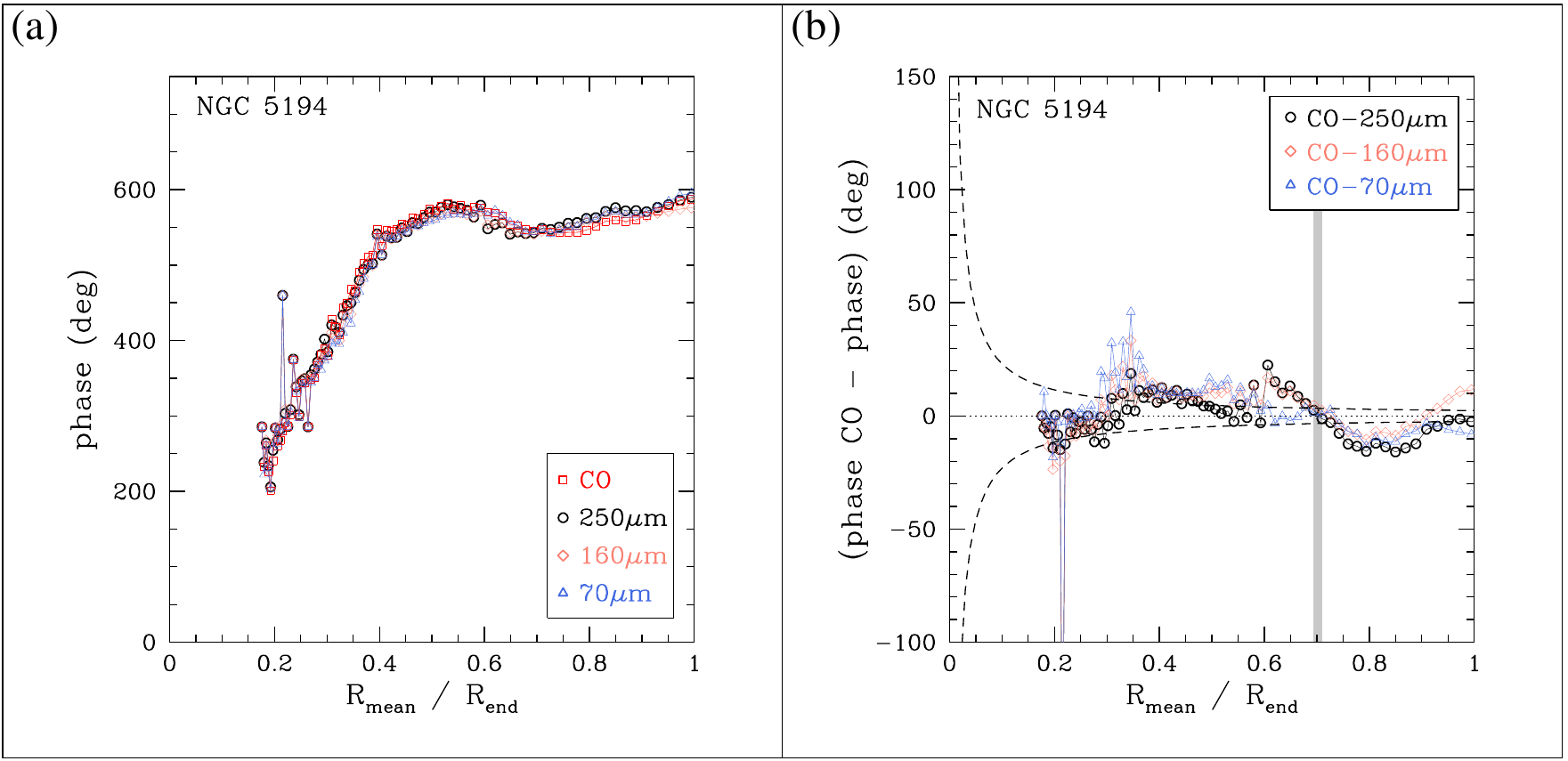}
\caption[f12]{Results for NGC~5194 (CO, $250\micron$, $160\micron$, \& $70\micron$).
{\it~Panel (a):} azimuthal phases in degrees ($y$-axis), vs.,
normalized radius $R_{\rm{mean}}/R_{\rm{end}}$ ($x$-axis).
Empty squares represent the CO,
empty circles represent the $250\micron$,
empty diamonds the $160\micron$,
and empty triangles the $70\micron$ emission.
{\it~Panel (b):} phase shifts between the phases in CO and other tracers ($y$-axis),
vs., normalized radius.
Empty circles represent the phase shift between CO and $250\micron$,
empty diamonds the phase shift between CO and $160\micron$,
and empty triangles the phase shift between CO and $70\micron$.
The gray shaded region shows the proposed corotation, $R_{\rm{mean}}/R_{\rm{end}}\sim0.7$, for the CO-$250\micron$ phase shifts.
The long-dashed line indicates the $\pm1-\sigma$ error.
~\label{N5194_COHerschel}}
\end{figure*}

\begin{figure*}
\centering
\includegraphics[scale=0.8,angle=0]{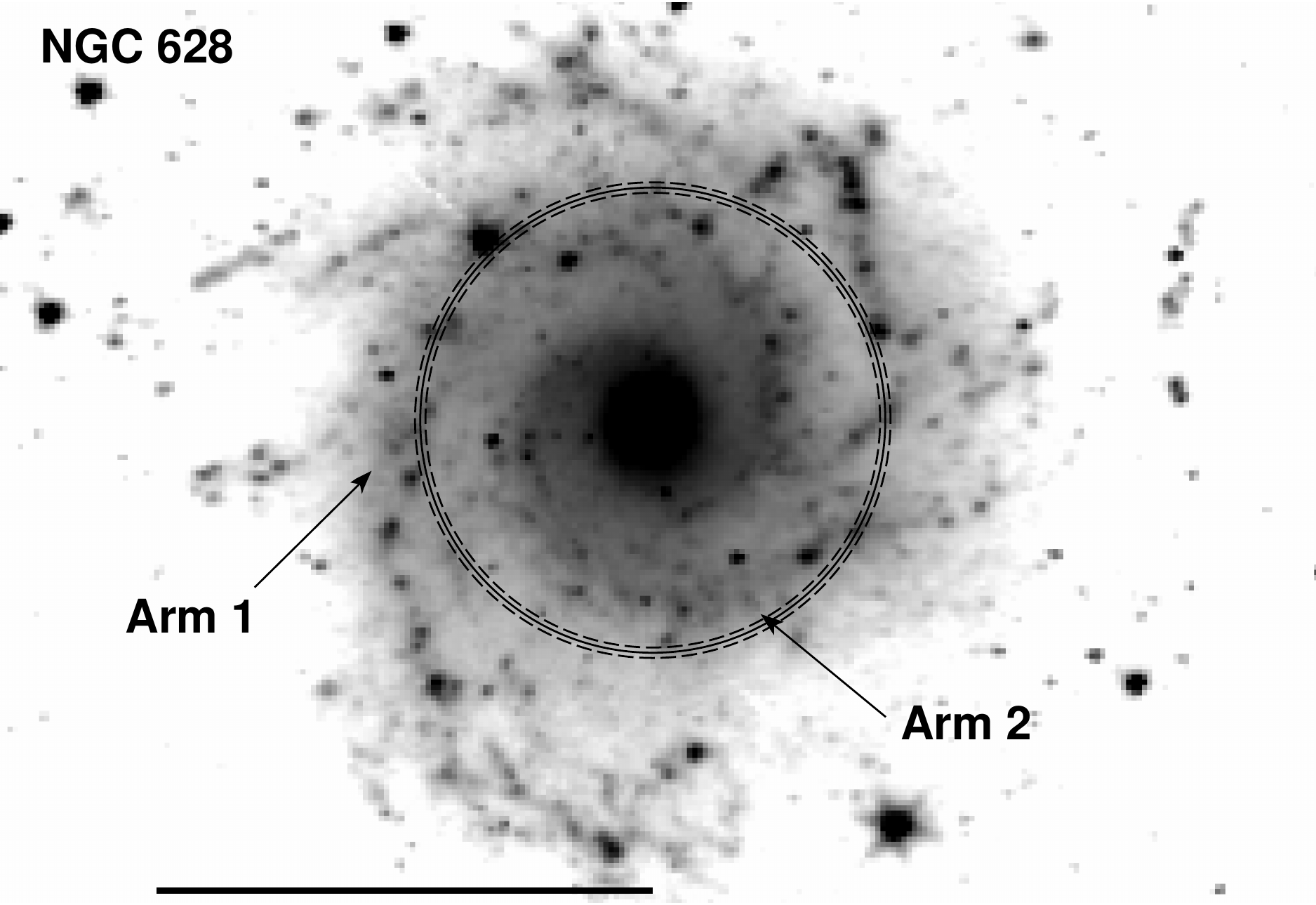}
\caption[f13.ps]{Nomenclature adopted to analyze separately the spiral arms of NGC~628.
The frame corresponds to the deprojected $3.6\mu$ image. Display is in logarithmic scale.
Dashed circles indicate the corotation region as inferred from the CO-$24\micron$ phase shifts
(see figure~\ref{N628_HICO24UV}),
continuous circle is the mean value~$\sim89\arcsec$ (see section~\ref{results}).
The horizontal line represents the maximum radial extent of the spiral arms,
$R_{\rm{end}}=190\arcsec$, measured from the galaxy's center (see also table~\ref{tbl-2}).
~\label{M74_arms}}
\end{figure*}

\begin{figure*}
\centering
\includegraphics[scale=0.8,angle=0]{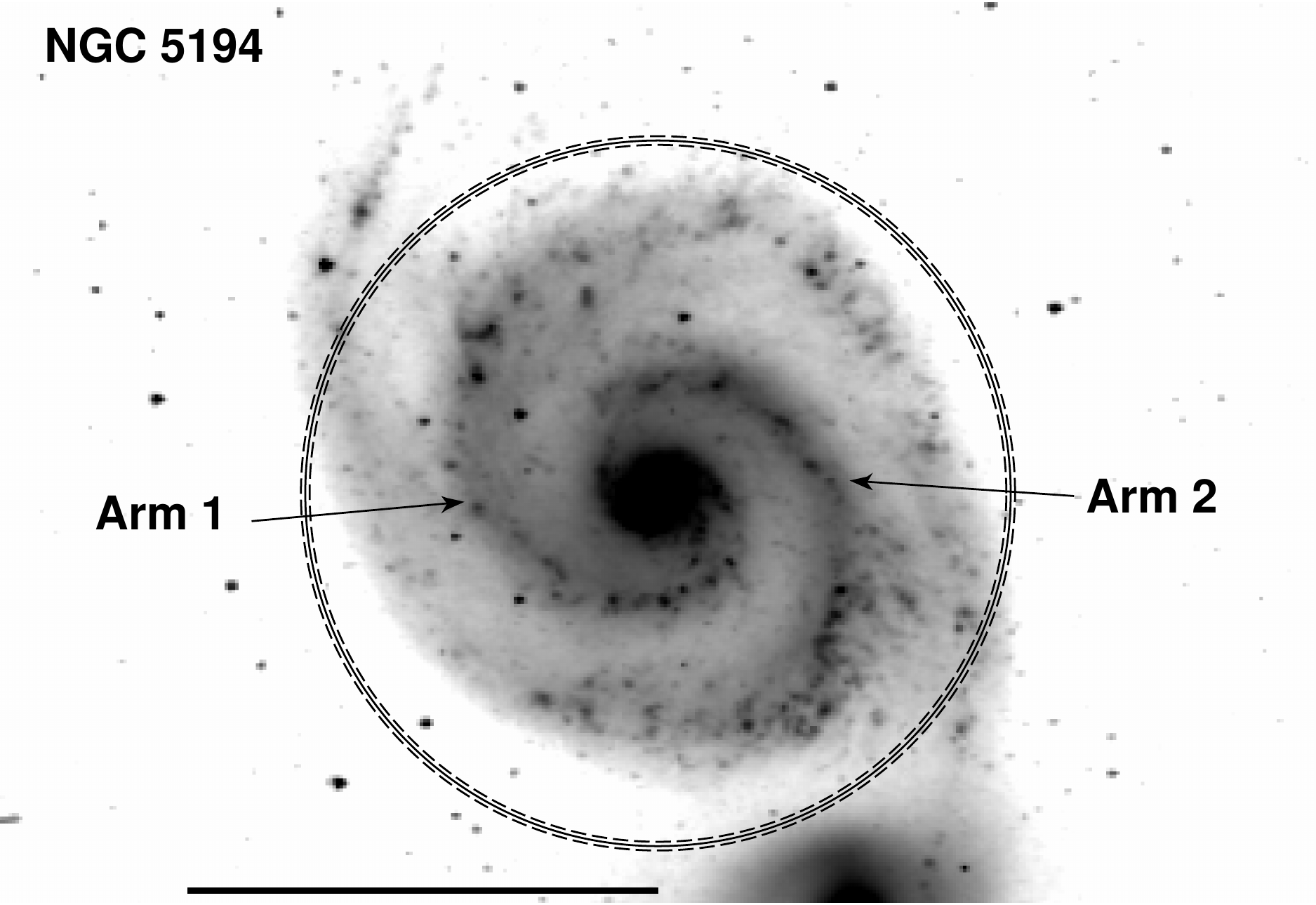}
\caption[f14.ps]{Nomenclature adopted to analyze separately the spiral arms of NGC~5194.
The frame corresponds to the deprojected $3.6\mu$ image. Display is in logarithmic scale.
Dashed circles indicate the corotation region as inferred from the CO-$24\micron$ phase shifts
(see figure~\ref{N5194_HICO24UV}),
continuous circle is the mean value~$\sim202\arcsec$ (see section~\ref{results}).
The horizontal line represents the maximum radial extent of the spiral arms,
$R_{\rm{end}}=270\arcsec$, measured from the galaxy's center (see also table~\ref{tbl-2}).
~\label{M51_arms}}
\end{figure*}

\begin{figure*}
\centering
\includegraphics[scale=0.8,angle=0]{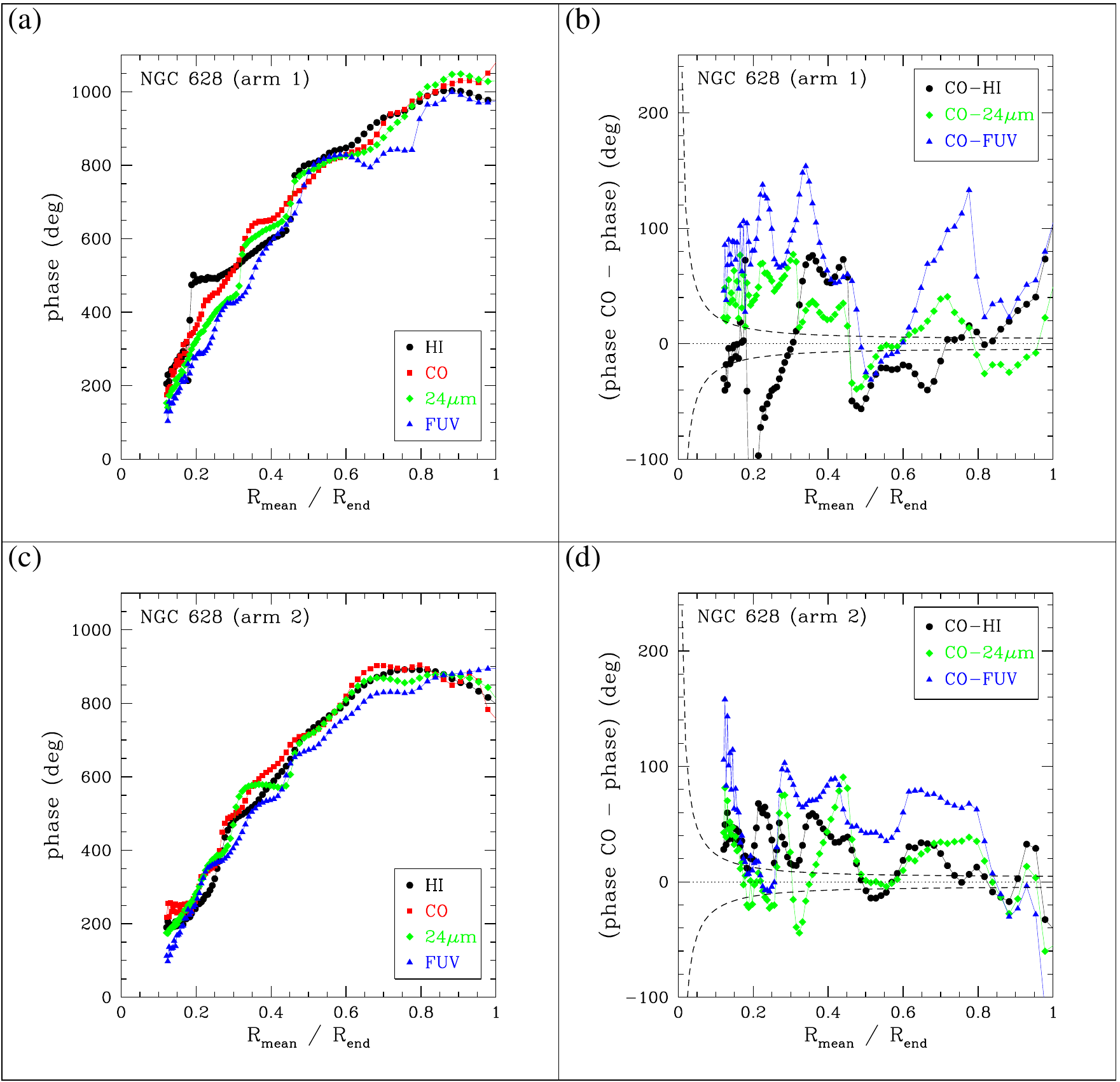}
\caption[f15]{
Mode $m=1$ results for NGC~628 (H{\rm{I}}, CO, $24\micron$, \& FUV).
{\it~Panels (a) and (c):} azimuthal phases in degrees ($y$-axis), vs.,
normalized radius $R_{\rm{mean}}/R_{\rm{end}}$ ($x$-axis).
{\it~Panels (b) and (d):} phase shift between CO and other tracers ($y$-axis),
vs., normalized radius.
Same symbolism as in figure~\ref{N628_HICO24UV}.
~\label{N628_isolarms}}
\end{figure*}

\begin{figure*}
\centering
\includegraphics[scale=0.8,angle=0]{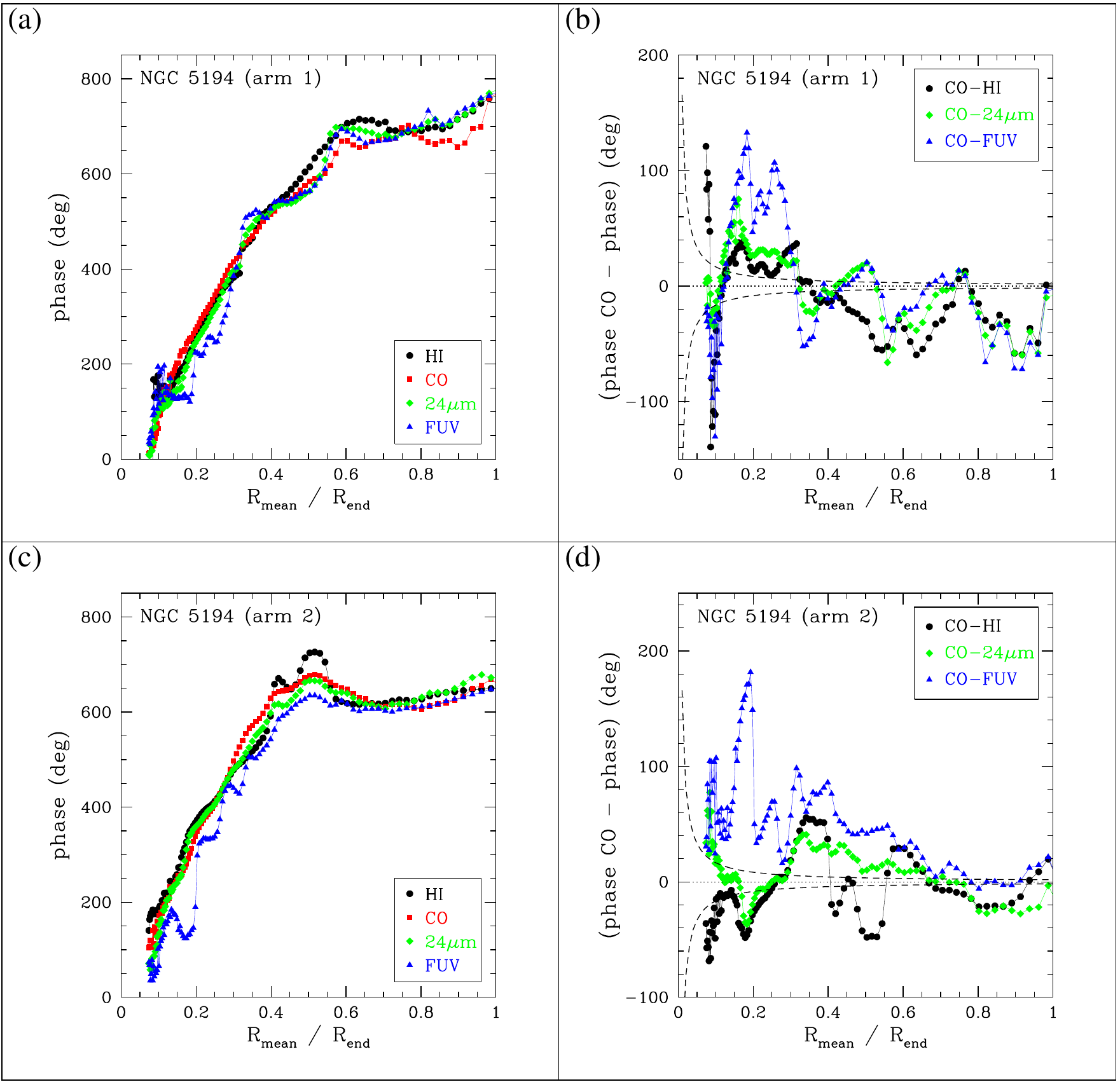}
\caption[f16]{
Mode $m=1$ results for NGC~5194 (H{\rm{I}}, CO, $24\micron$, \& FUV).
{\it~Panels (a) and (c):} azimuthal phases in degrees ($y$-axis), vs.,
normalized radius $R_{\rm{mean}}/R_{\rm{end}}$ ($x$-axis).
{\it~Panels (b) and (d):} phase shift between CO and other tracers ($y$-axis),
vs., normalized radius.
Same symbolism as in figure~\ref{N628_HICO24UV}.
~\label{N5194_isolarms}}
\end{figure*}

\begin{figure*}
\centering
\includegraphics[scale=1.0,angle=0]{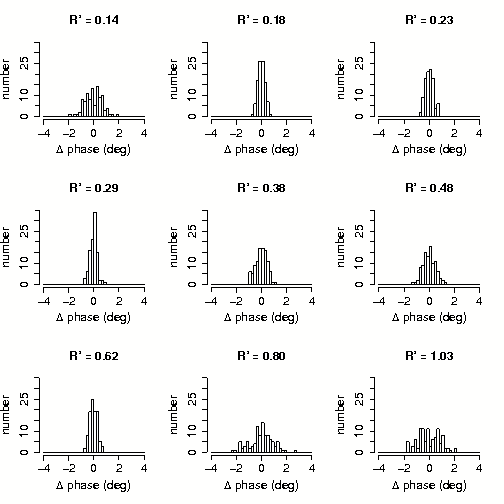}
\caption[f17]{Probability distributions of the Fourier phases for different radii
($R'=R_{\rm{mean}}/R_{\rm{end}}$), obtained with Monte Carlo methods for the CO
integrated intensity map of NGC~5194~\citep{ler09}.
The $\Delta$ phase values are shown in degrees.
~\label{phases_prob_M51}}
\end{figure*}

\begin{figure*}
\centering
\includegraphics[scale=0.6,angle=0]{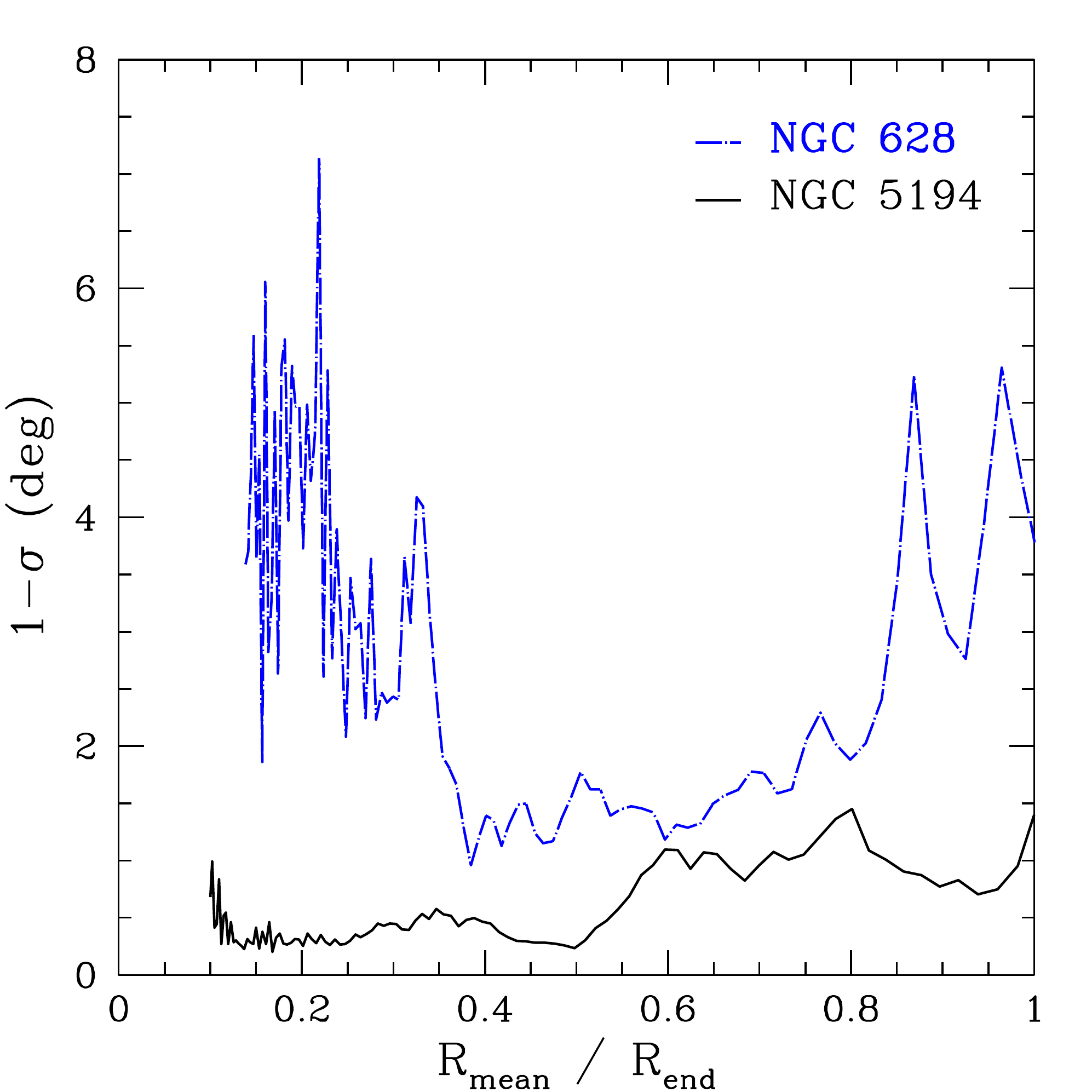}
\caption[f18]{Uncertainties, $1-\sigma$ ($y$-axis, in degrees), of the Fourier phases according
to the Monte Carlo simulations outcome, vs., the normalized radius $R_{\rm{mean}}/R_{\rm{end}}$ ($x$-axis).
The dashed-dotted line indicates the uncertainties for NGC~628, with a median value of $3.0\degr$, while the
continuous line indicate the uncertainties for NGC~5194, with a median value of $0.5\degr$.
~\label{one_sigma_err}}
\end{figure*}

\clearpage

\begin{figure*}
\centering
\includegraphics[scale=0.8,angle=0]{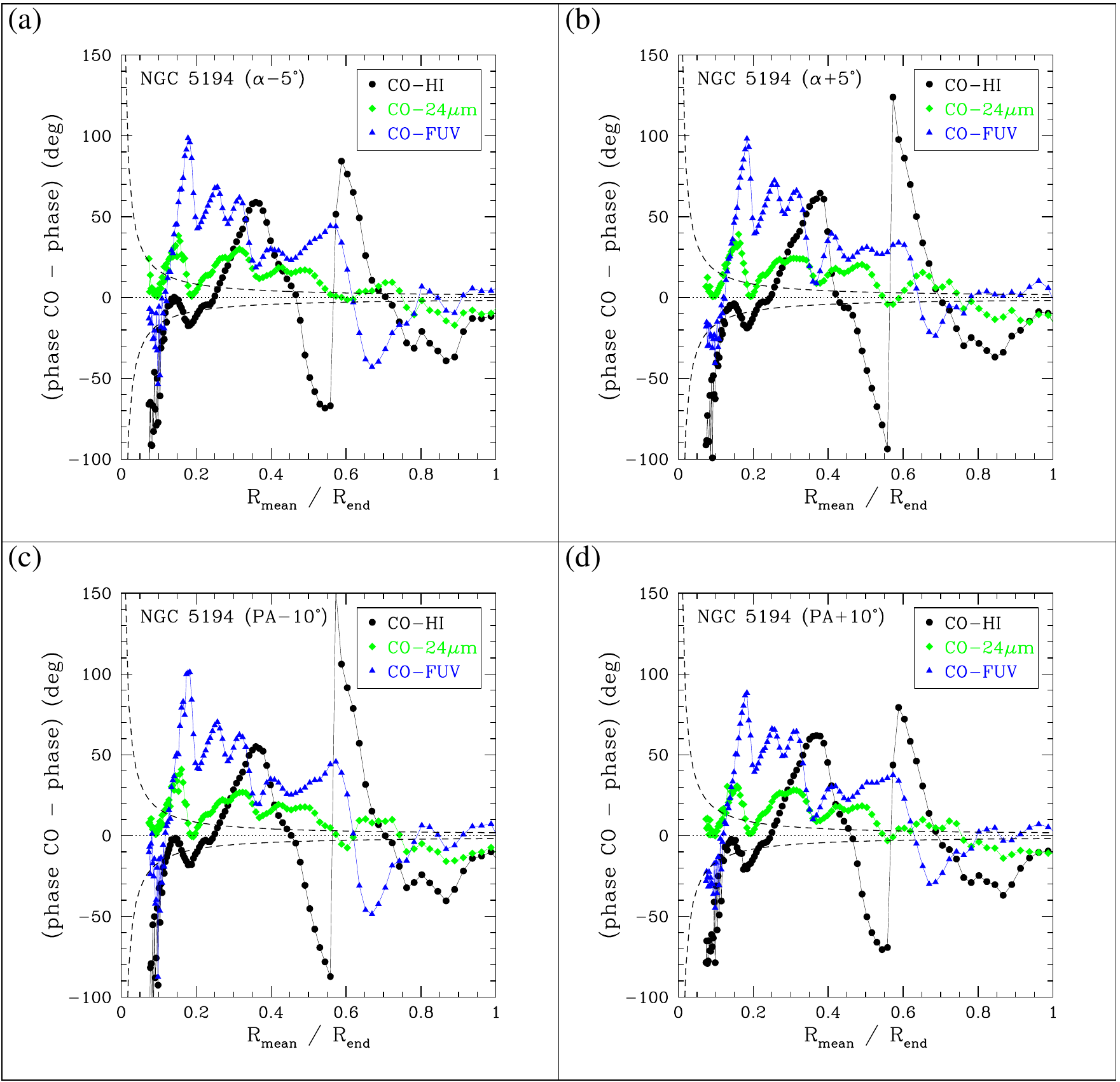}
\caption[f19]{Results (for H{\rm{I}}, CO, $24\micron$, \& FUV) obtained by varying the projection parameters of NGC~5194.
{\it~Panels (a) and (b):} the inclination angle, $\alpha$, was varied by $\pm5\degr$.
{\it~Panels (c) and (d):} the position angle, PA, was varied by $\pm10\degr$.
Same symbolism as in figure~\ref{N628_HICO24UV}.
~\label{proyec_param}}
\end{figure*}


\end{document}